\documentclass[conference]{IEEEtran}
\IEEEoverridecommandlockouts
\usepackage{cite}
\usepackage{amsmath,amssymb,amsfonts}
\usepackage{algorithmic}
\usepackage{graphicx}
\usepackage{textcomp}
\usepackage{xcolor}
\def\BibTeX{{\rm B\kern-.05em{\sc i\kern-.025em b}\kern-.08em
    T\kern-.1667em\lower.7ex\hbox{E}\kern-.125emX}}
    
\usepackage{subfiles}
\usepackage{paralist}
\usepackage{amsmath,amsfonts,amsthm}
\usepackage{mathrsfs}
\usepackage{amsbsy}
\usepackage[mathscr]{eucal} 
\usepackage{algorithmic}

\usepackage[noend,ruled,vlined,linesnumbered,resetcount]{algorithm2e}
\usepackage{booktabs}
\usepackage{graphicx}
\usepackage[T1]{fontenc}
\usepackage{textcomp}
\usepackage{xcolor}
\usepackage{comment}
\usepackage[hidelinks]{hyperref}

\usepackage{multirow}
\usepackage{tabularx,stackengine,collcell}
\let\endminwd\relax
\newcolumntype{L}[1]{>{\collectcell\xminwd l{#1}}l<{\endminwd\endcollectcell}}
\newcolumntype{C}[1]{>{\collectcell\xminwd c{#1}}c<{\endminwd\endcollectcell}}
\newcolumntype{R}[1]{>{\collectcell\xminwd r{#1}}r<{\endminwd\endcollectcell}}
\def\minwd#1#2#3\endminwd{\stackengine{0pt}{#3}{\rule{#2}{0pt}}{O}{#1}{F}{F}{L}}
\newcommand\xminwd[1]{\minwd#1}

\usepackage{subcaption}
\usepackage[font=normalsize]{caption}
\usepackage{bbold}
\usepackage{mathtools}
\usepackage{enumitem}
\usepackage{bbm}

\newtheorem{lemma}{Lemma}

\SetAlFnt{\small}
\SetKwComment{Comment}{$\triangleright$\ }{}

\SetCommentSty{mycommfont}
\SetKwInput{KwData}{Input}
\SetKwInput{KwResult}{Output}
\newcommand{\method}{\textsc{MARIOH}\xspace}

\newcommand{\wang}{\textsc{SHyRe}\xspace}

\newcommand{\wangunsup}{\textsc{SHyRe-Unsup}\xspace}

\newcommand{\demon}{\textsc{Demon}\xspace}

\newcommand{\cfinder}{\textsc{CFinder}\xspace}

\newcommand{\ecc}{\textsc{Clique Covering}\xspace}

\newcommand{\maxclique}{\textsc{Max Clique}\xspace}

\newcommand{\bayesian}{\textsc{Bayesian-MDL}\xspace}

\newtheoremstyle{problemstyle}  
{3pt}                                               
{3pt}                                               
{\normalfont}                               
{}                                                  
{\bfseries\itshape}                 
{\normalfont\bfseries:}         
{.5em}                                          
{}                                                  
\theoremstyle{problemstyle}

\newtheorem{problem}{Problem}

\newcommand{\smallsection}[1]{{\vspace{0.05in} \noindent {\bf{\underline{\smash{#1}}}}}}

\def\endthebibliography{%
	\def\@noitemerr{\@latex@warning{Empty `thebibliography' environment}}%
	\endlist
}

\begin{document}

\title{MARIOH: Multiplicity-Aware Hypergraph Reconstruction}
\author{
\IEEEauthorblockN{Kyuhan Lee\textsuperscript{1,2}, Geon Lee\textsuperscript{1}, and Kijung Shin\textsuperscript{1}}
\IEEEauthorblockA{\textsuperscript{1}Kim Jaechul Graduate School of AI, KAIST, Seoul, Republic of Korea,
\textsuperscript{2}GraphAI, Daejeon, Republic of Korea
\\
\{kyuhan.lee, geonlee0325, kijungs\}@kaist.ac.kr} 
}


\maketitle

\begin{abstract} 
Hypergraphs offer a powerful framework for modeling higher-order interactions that traditional pairwise graphs cannot fully capture. However, practical constraints often lead to their simplification into projected graphs, resulting in substantial information loss and ambiguity in representing higher-order relationships. In this work, we propose \method, a supervised approach for reconstructing the original hypergraph from its projected graph by leveraging edge multiplicity.
To overcome the difficulties posed by the large search space,
\method integrates several key ideas: (a) identifying provable size-$2$ hyperedges, which reduces the candidate search space, (b) predicting the likelihood of candidates being hyperedges by utilizing both structural and multiplicity-related features, and (c) not only targeting promising hyperedge candidates but also examining less confident ones to explore alternative possibilities.
Together, these ideas enable \method to efficiently and effectively explore the search space.
In our experiments using 10 real-world datasets, \method achieves up to 74.51\% higher reconstruction accuracy compared to state-of-the-art methods.


\end{abstract}


\section{Introduction}
\label{sec:intro}
Hypergraphs extend ordinary graphs by allowing edges, known as hyperedges, to connect more than two nodes. This feature makes hypergraphs useful for modeling higher-order interactions involving multiple entities~\cite{battiston2020networks, torres2021and}. High-order interactions occur in various systems, including co-authorship networks, where multiple authors collaborate on a paper~\cite{benson2018simplicial}, email networks, where an email is sent to multiple recipients~\cite{benson2018simplicial}, 
and social networks with group interactions~\cite{yang2019revisiting, antelmi2021social}.

\begin{figure}[t]
    \centering
    \includegraphics[width=\linewidth]{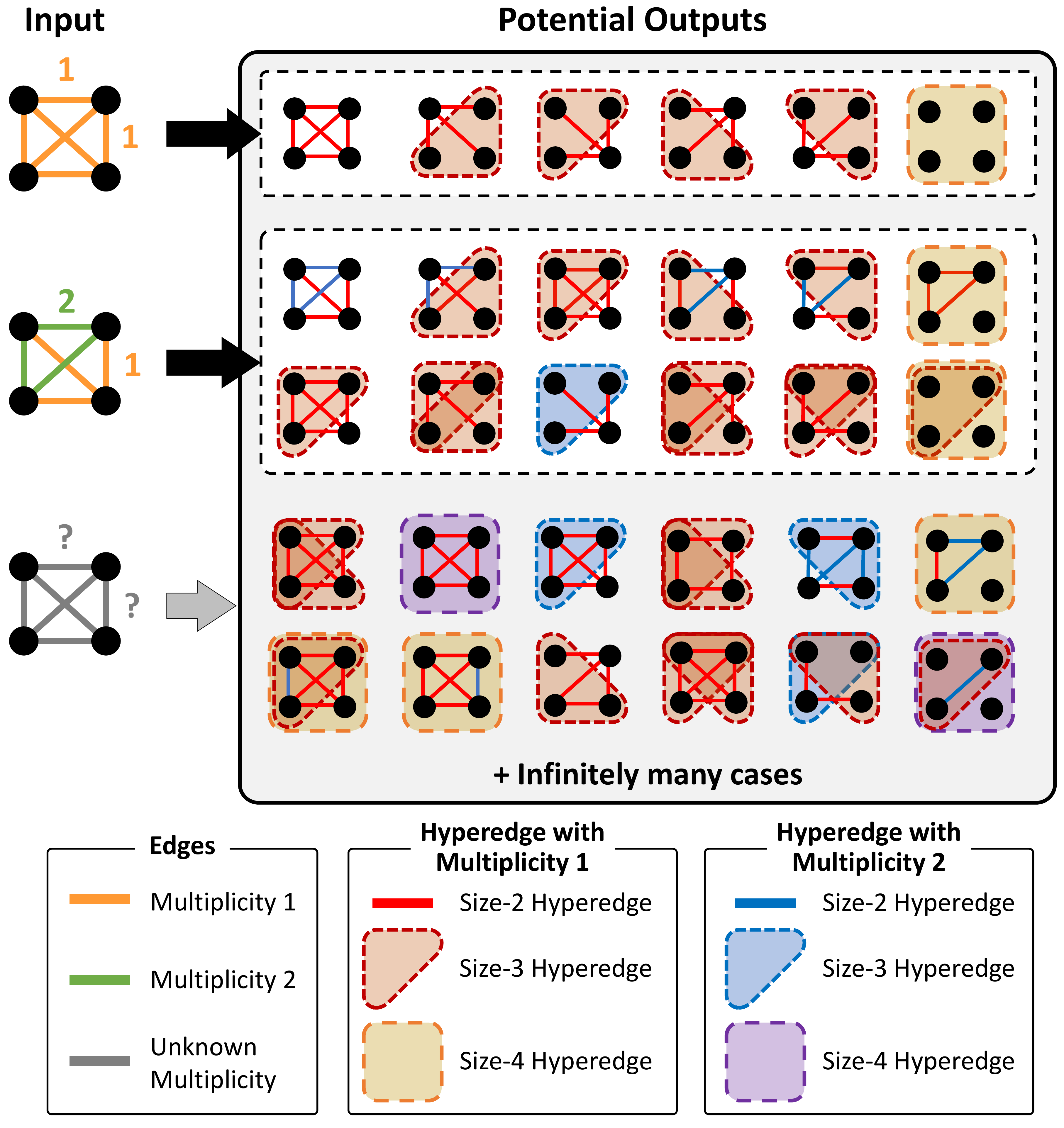}
    \caption{\small{\label{fig:candidate_space} 
        \underline{\smash{Reduction of candidate space using edge multiplicity.}} When edge multiplicities are known (top and middle rows), the number of potential outputs is significantly reduced compared to cases with unknown multiplicity (bottom row). The known edge multiplicities (e.g., multiplicity 1 and 2) constrain the possible hyperedge structures, limiting the search space and enabling more accurate reconstruction. In contrast, the absence of edge multiplicity information leads to an explosion of candidates, including infinitely many possibilities, complicating the reconstruction process.}}


\end{figure}


\begin{figure*}[t]
	\vspace{-3mm}
	\centering
	\includegraphics[width= \linewidth]{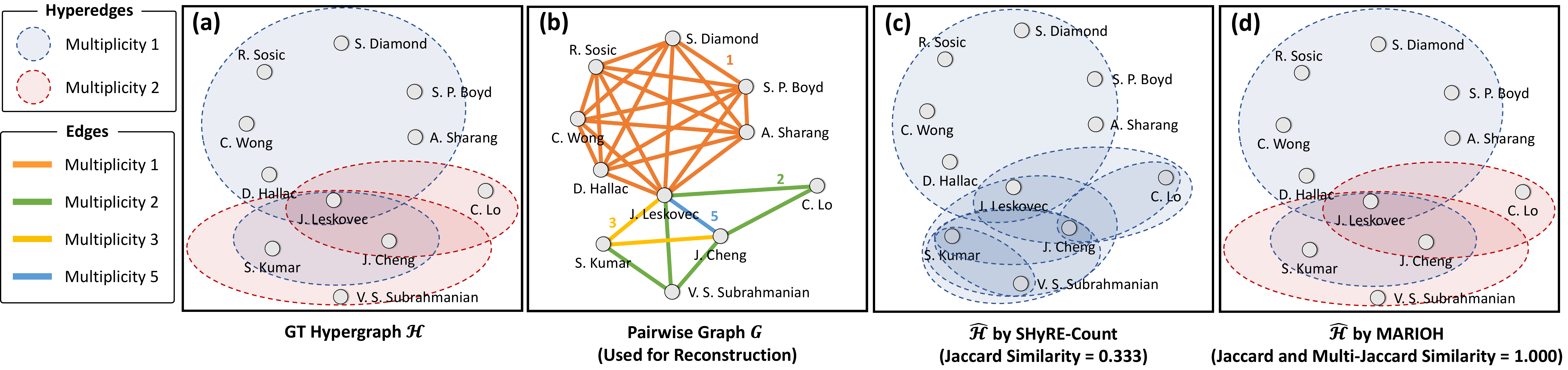}
        \vspace{-6mm}
	\caption{\label{fig:case_study}
		\small{\underline{\smash{Example of hypergraph reconstruction on a co-authorship dataset.}}
        \textbf{(a):}
        In the ground-truth hypergraph, each hyperedge represents a set of researchers who co-authored a paper. We focus on the visualized sub-hypergraph $\mathcal{H}$, induced by Jure Leskovec and his randomly chosen ten co-authors.
        \textbf{(b):} As input, the graph representation $G$ of $\mathcal{H}$ is given, where each edge indicates the number of co-authored papers between researchers.
        \textbf{(c):} \wang-Count \cite{wang2024from}  recovers a subset of ground-truth hyperedges in $\mathcal{H}$ along with some false positives
        (Jaccard similarity $= 0.333$).
        \textbf{(d):} In contrast, our proposed method, \method, exactly restores $\mathcal{H}$
        (Jaccard and multi-Jaccard similarity $= 1.000$).
        Refer to Section~\ref{exp:case} for detailed settings, and refer to the online appendix \cite{supple}
        for more case studies on the Host-virus and Crimes datasets.}
	}
\end{figure*}

Although hypergraphs offer advantages in modeling higher-order interactions, practical challenges often lead to their simplification into pairwise graphs. Examples include: 
\begin{itemize}[leftmargin=*]
    \item \textbf{Biological Complexes:} In cellular systems, multiple proteins and biomolecules interact to form complexes or signal transduction assemblies. However, standard assays, such as the yeast two-hybrid system~\cite{fields1989novel} or tandem affinity purification~\cite{gavin2006proteome, krogan2006global}, mainly capture only pairwise interactions. 
    \item \textbf{Social Networks:} People interact in groups in meetings and collaborative projects, but available social network data usually record only pairwise interactions~\cite{resnick1997protecting, eagle2006reality, young2021hypergraph}. 
    Since group interactions are often brief, context-dependent, and subject to privacy issues, it is hard to observe them directly. 
    \item \textbf{Brain Connectivity:} The brain functions through the coordinated activity of multiple regions at once. However, neuroimaging techniques, such as fMRI, typically provide only pairwise correlation measures between brain regions~\cite{xiao2019multi}. 
\end{itemize}
As a result, the underlying hypergraph datasets are often unavailable~\cite{newman2004coauthorship, sarigol2014predicting}, with only their simplified graph representations being released~\cite{leskovec2005graphs, hu2020open}.

Hypergraph reconstruction, illustrated in Fig.~\ref{fig:case_study}, aims to recover the original hypergraph from its graph-based representation, addressing the loss of information from simplification. Simplifying hypergraphs into graphs results in information loss and ambiguity, as different higher-order interactions can share the same pairwise-edge representations\cite{zhou2006learning}. For example, a size-3 clique in the graph could represent a single hyperedge involving three nodes or three separate pairwise hyperedges, making it challenging to determine the original hyperedge structure. 
Reconstruction is further complicated by the large search space—any clique in the graph could potentially correspond to a hyperedge, and the number of cliques is typically orders of magnitude greater than the number of nodes or edges.
This complexity grows when the same hyperedge may appear multiple times, as we must determine not only which groups of nodes form hyperedges but also their frequency.


Despite its difficulty, hypergraph reconstruction is crucial for many reasons. First, it improves \textbf{applicability} by enabling the use of hypergraph-based tools, which often outperform their graph-only counterparts in various tasks, including clustering\cite{zhou2006learning}, node classification\cite{feng2019hypergraph, dong2020hnhn, chien2022you, huang2021unignn}, link prediction\cite{yoon2020much}, and anomaly detection\cite{lee2022hash}
(see Section~\ref{sec:exp:application} for our experiments on applicability). Second, it offers \textbf{storage efficiency}. While representing a clique of size $N$ in a graph requires storing $\binom{N}{2}$ edges, a hypergraph representation needs only $O(N)$ space for the corresponding hyperedge (see the online appendix\cite{supple} for detailed storage savings).
Third, it enables a \textbf{deeper understanding of underlying systems} (e.g., brains) by uncovering higher-order relationships (e.g., multi-region interactions) that are lost in graph representations but crucial for interpreting individual or system-level behaviors~\cite{sporns2014contributions, tang2023comprehensive}.
Existing methods for hypergraph reconstruction include a Bayesian approach proposed by Young et al.\cite{young2021hypergraph}, which seeks to minimize the number of cliques used to cover the projected graph based on the principle of parsimony. However, this principle does not always hold in practice\cite{wang2024from}.
Wang and Kleinberg\cite{wang2024from} introduced \wang, a supervised method that samples cliques from the projected graph and classifies them as either hyperedges or non-hyperedges. While this method leverages machine learning for classification, it relies on sampling, which introduces the issue of false negatives—potential hyperedges not sampled are missed entirely. 

Importantly, these methods often disregard edge multiplicity in the given graph, even though it can significantly reduce the search space. Edge multiplicity indicates how many times a particular pair of nodes (i.e., an edge in the projected graph) co-occurs across different hyperedges. Notably, this differs from hyperedge multiplicity, which indicates how many times an identical hyperedge (the same set of nodes) is repeated within the hypergraph. As illustrated in Fig.~\ref{fig:candidate_space}, when edge multiplicity information is absent, it is unclear how many times higher-order interactions containing each edge occurred (only a minimum of once is guaranteed). This ambiguity causes an explosion in the search space, especially in duplicated cases, where multiple higher-order interactions share the same set of nodes. For instance, a size-4 clique composed entirely of unknown edges can represent an infinite number of possible higher-order interactions, as the exact count for each edge is unknown.
In contrast, when edge multiplicity information is available, it significantly reduces the candidate space for duplicated cases, as shown in Fig.~\ref{fig:candidate_space}. Knowing how many times each edge participates in higher-order interactions limits the possible outputs, enabling more accurate reconstruction.
To the best of our knowledge, the only discussion on leveraging multiplicity can be found in the appendix of \cite{wang2024from}. With edge multiplicity in mind, the authors proposed an unsupervised approach where, at each iteration, the highest-ranked maximal clique—determined by specific conditions—is replaced by the corresponding hyperedge. While this method considers edge multiplicity, 
it suffers from scalability issues on large datasets, since maximal cliques are replaced one by one, requiring the repetition of computationally expensive searches of them. 
Moreover, in some cases, its reconstruction accuracy is even worse than \wang, which does not utilize edge multiplicity (as shown in\cite{wang2024from}), highlighting its limitations.

Driven by these limitations, in this paper, we propose \method, a supervised method for multiplicity-aware hypergraph reconstruction. \method starts with the projected graph and employs a multiplicity-aware classifier to assign prediction scores to potential hyperedge candidates, including maximal cliques and their subsets. It then greedily replaces the highest-scoring candidates with the corresponding hyperedges.

\method has three key components. First, we reduce the search space by identifying size-$2$ hyperedge candidates that are theoretically guaranteed to be true hyperedges, minimizing false positives. Second, we use a classifier trained on not just structural but also multiplicity-related features from the projected graph to predict the likelihood of each hyperedge candidate being a true hyperedge, improving selection by accounting for multiplicity. 
Finally, \method employs a bidirectional search that examines both the most promising and the least promising hyperedge candidates (i.e., cliques). By not just focusing on highly promising maximal cliques but also examining sub-cliques contained in less promising ones, we ensure that important hyperedges are not overlooked. 


Through extensive experiments on 10 real-world datasets, we demonstrate the superiority of \method over eight baseline approaches. We summarize its strengths as follows:
\begin{itemize}[leftmargin=*]
    \item \textbf{Accuracy:}
    \method recovers real-world hypergraphs up to 74.51\% more accurately than baseline methods.
    \item \textbf{Transferability:} 
    \method trained on a hypergraph is applied effectively to (the projected graph of) another hypergraph if they are from the same domain (e.g., two co-authorship networks from different fields of study).
    
    \item \textbf{Applicability:} Compared to using a projected graph, using the hypergraph reconstructed from it by \method improves performance in downstream tasks, including link prediction and node clustering. 
\end{itemize}

\smallsection{Reproducibility:} The code and data used in the paper can be found at \url{https://github.com/KyuhanLee/MARIOH}.

	
\section{Preliminaries and Problem Definition}
\label{sec:preliminaries}
\subsection{\bf Preliminaries}
Let \( \mathcal{V} \) be a set of nodes. A \textit{hypergraph} is represented as a pair \( \mathcal{H} = (\mathcal{V}, \mathcal{E}_\mathcal{H}^{*}) \), where \( \mathcal{E}_\mathcal{H}^{*} \) is the multiset of hyperedges. Each hyperedge \( e \in \mathcal{E}_\mathcal{H}^{*} \) is a subset of \( \mathcal{V} \) with at least two nodes, i.e., \( e \subseteq \mathcal{V} \) and \( |e| \geq 2 \).
That is, multiple hyperedges comprising the same set of nodes can exist in the multiset $\mathcal{E}_\mathcal{H}^{*}$.
We define the \textit{multiplicity} $M_{\mathcal{H}}(e)$ of a hyperedge $e\in\mathcal{E}_\mathcal{H}$ as the number of times $e$ appears in $\mathcal{E}_\mathcal{H}^{*}$, i.e., $M_{\mathcal{H}}(e)= | \{e'\in \mathcal{E}_\mathcal{H}^{*} : e' = e\} |$.
Equivalently, the hypergraph can be represented as $\mathcal{H}=(\mathcal{V}, \mathcal{E}_\mathcal{H}, M_{\mathcal{H}})$, where $\mathcal{E}_\mathcal{H}$ is the set of unique hyperedges in the multiset $\mathcal{E}_\mathcal{H}^{*}$, and $M_{\mathcal{H}}: \mathcal{E}_\mathcal{H}^{*} \rightarrow \mathbb{N}$ is a function that assigns the multiplicity of each hyperedge.

\textit{Clique expansion} of a hypergraph \( \mathcal{H} \) results in its (weighted) projected graph \( \mathcal{G} = (\mathcal{V}, \mathcal{E}_\mathcal{G}, \omega) \), where \( \mathcal{E}_\mathcal{G} \) is the set of node pairs that co-appear in at least one hyperedge in $\mathcal{E}_\mathcal{H}$:
\begin{equation*}
\mathcal{E}_\mathcal{G} = \{\{u,v\} : \exists e\in \mathcal{E}_\mathcal{H} \ \text{s.t.}\ \{u,v\}\subseteq e\}.    
\end{equation*}
The weight $\omega_{u,v}$ of an edge $\{u,v\}$ corresponds to the number of hyperedges in $\mathcal{H}$ that contain both nodes:
\begin{equation*}
    \omega_{u,v} = \sum_{e\in \mathcal{E}_\mathcal{H}^{*}} \mathbb{1}(\{u, v\}\subseteq e) = \sum_{e\in \mathcal{E}_\mathcal{H}} M_\mathcal{H}(e)\cdot \mathbb{1}(\{u,v\}\subseteq e),
\end{equation*}
\noindent
where $\mathbb{1}(\cdot)$ is the indicator function.


A \textit{clique} in the projected graph $\mathcal{G}=(\mathcal{V},\mathcal{E}_\mathcal{G}, \omega)$ is a subset $\mathcal{C} \subseteq \mathcal{V}$ such that every pair of distinct nodes in $\mathcal{C}$ is connected by an edge, i.e., $\forall u,v\in \mathcal{C}, u\neq v \implies \{u,v\}\in \mathcal{E}_\mathcal{G}$.
That is, a subset of nodes $\mathcal{C}$ forms a clique if and only if the subgraph of $\mathcal{G}$ induced by $\mathcal{C}$ is a complete graph. 
A \textit{maximal clique} is a clique that is not a subset of any larger clique in $\mathcal{G}$.

\begin{figure*}[t]
	\vspace{-3mm}
	\centering
	\includegraphics[width= \linewidth]{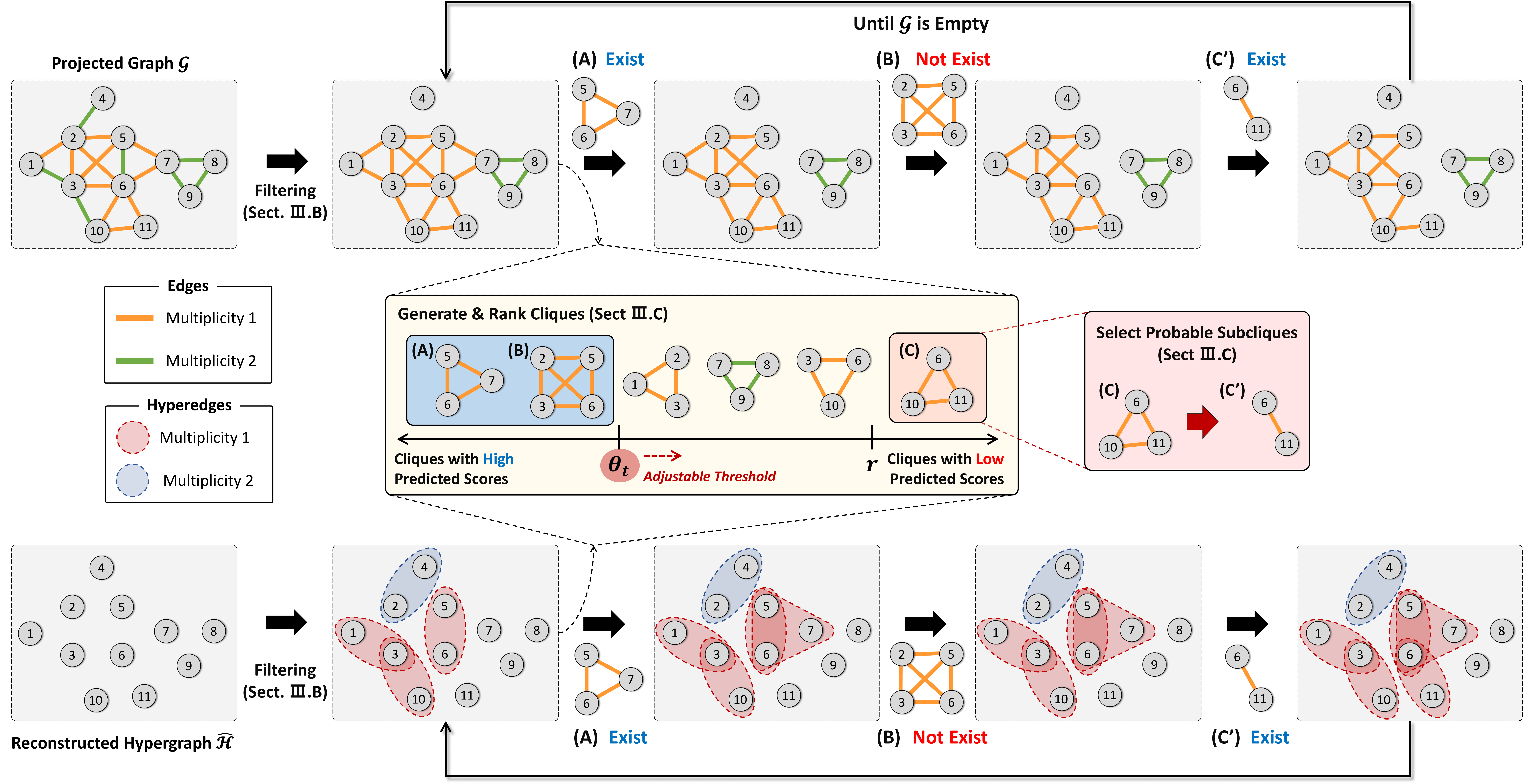}
	\caption{\label{fig:model}
		\small{\underline{\smash{Example procedure of \method.}}
        From a projected graph $\mathcal{G}$ (top), \method reconstructs a hypergraph $\widehat{\mathcal{H}}$ (bottom).
        First, \method identifies edges in $\mathcal{G}$ that are theoretically guaranteed to correspond to size-$2$ hyperedges.
        These edges are directly incorporated into $\widehat{\mathcal{H}}$ and removed from $\mathcal{G}$.
        Then, \method identifies the maximal cliques in $\mathcal{G}$ and predicts the likelihood of each clique using a classifier trained on multiplicity-aware features.
        Based on the predicted likelihood, \method employs a bidirectional search that identifies (1) cliques with high estimated likelihood (e.g., (A) $\{5,6,7\}$ and (B) $\{2,3,5,6\}$) and (2) sub-cliques with high likelihood (e.g., (C') $\{6,11\}$) but are hidden within larger low-likelihood cliques (e.g., (C) $\{6,10,11\}$), as hyperedges.
        However, clique (B) $\{2,3,5,6\}$ is not identified as a hyperedge since it no longer exists in the updated projected graph after removing (A) $\{5,6,7\}$.
        This is repeated until no edges remain in $\mathcal{G}$.
	}}
\end{figure*}

\subsection{\bf Problem Definition}

In this work, we generally follow the supervised problem setting in\cite{wang2024from} (a detailed comparison between the settings is provided below). 
As discussed in
Sect.~\ref{sec:intro}, reconstructing a hypergraph from its projected graph is extremely difficult due to the loss of higher-order relationships. 
However, if available, a hypergraph from the same domain provides valuable domain-specific information for the problem, since each domain has unique structural patterns and characteristics\cite{lee2020hypergraph, lotito2022higher, larock2023encapsulation}.
Thus, Wang and Kleinberg\cite{wang2024from} assume the existence of such a hypergraph, as part of the input, for supervision, leading to the following problem definition: 
\begin{problem}[\label{problem}Supervised Hypergraph Reconstruction]~
\begin{itemize}[leftmargin=*]
    \item \textbf{Given}: 
    Source and target (hyper)graph information:
    \begin{itemize}[leftmargin=*]
        \item a \textit{source} projected graph \( \mathcal{G}^{(S)} = (\mathcal{V}^{(S)}, \mathcal{E}_{\mathcal{G}}^{(S)}, \omega^{(S)}) \)
        \item a corresponding \textit{source} hypergraph \( \mathcal{H}^{(S)} = (\mathcal{V}^{(S)}, {\mathcal{E}_\mathcal{H}^{*(S)}}) \)
        \item a \textit{target} projected graph \(\mathcal{G}^{(T)} = (\mathcal{V}^{(T)}, \mathcal{E}_\mathcal{G}^{(T)}, \omega^{(T)}) \)
    \end{itemize}
    \item \textbf{Find}: a reconstructed hypergraph \( \widehat{\mathcal{H}}^{(T)} = (\mathcal{V}^{(T)}, \widehat{\mathcal{E}}_\mathcal{H}^{*(T)}) \) corresponding to the \textit{target} projected graph $\mathcal{G}^{(T)}$
    \item \textbf{to Maximize}: the reconstruction accuracy.
\end{itemize}
\end{problem}

\noindent
Note that the source and target (hyper)graphs are from the same domain. 
Learning to reconstruct the source projected graph $\mathcal{G}^{(S)}$ into the original source hypergraph $\mathcal{H}^{(S)}$ provides valuable guidance for accurately reconstructing the target projected graph $\mathcal{G}^{(T)}$ into a target hypergraph $\mathcal{H}^{(T)}$.

\smallsection{Reconstruction Accuracy:}
To numerically evaluate the similarity between the reconstructed hypergraph $\widehat{\mathcal{H}}^{(T)}$ and the original target hypergraph $\mathcal{H}^{(T)}$, we measure the Jaccard similarity, following previous studies on hypergraph reconstruction~\cite{young2021hypergraph, wang2024from}.
The Jaccard similarity $\mathcal{J}(\mathcal{H}^{(T)}, \widehat{\mathcal{H}}^{(T)})$ between two hypergraphs $\mathcal{H}^{(T)}$ and $\widehat{\mathcal{H}}^{(T)}$ is defined as:
\begin{equation*}
    \mathcal{J}(\mathcal{H}^{(T)}, \widehat{\mathcal{H}}^{(T)}) = \frac{|\mathcal{E}_\mathcal{H}^{(T)} \cap \widehat{\mathcal{E}}_\mathcal{H}^{(T)}|}{|\mathcal{E}_\mathcal{H}^{(T)} \cup \widehat{\mathcal{E}}_\mathcal{H}^{(T)}|}
\end{equation*}

\noindent 
where it evaluates how accurately the unique hyperedges are reconstructed in $\widehat{\mathcal{H}}^{(T)}$.

In addition, to account for the multiplicity of the hyperedges, we use the multi-Jaccard similarity\cite{Costa2021FurtherGO}, which extends the Jaccard similarity by incorporating the multiplicities of hyperedges. 
The multi-Jaccard similarity $\mathcal{J}_{\text{multi}}(\mathcal{H}^{(T)}, \widehat{\mathcal{H}}^{(T)})$ between $\mathcal{H}^{(T)}$ and $\widehat{\mathcal{H}}^{(T)}$ is defined as:

\begin{equation*}
\small
    \mathcal{J}_{\text{multi}}(\mathcal{H}^{(T)}, \widehat{\mathcal{H}}^{(T)}) = 
    \frac{\sum_{e\in \mathcal{E}_\text{union}} \min(M_{\mathcal{H}^{(T)}}(e), M_{\widehat{\mathcal{H}}^{(T)}}(e))}{\sum_{e\in \mathcal{E}_\text{union}} \max(M_{\mathcal{H}^{(T)}}(e), M_{\widehat{\mathcal{H}}^{(T)}}(e))}
\end{equation*}

\noindent 
where $\mathcal{E}_\text{union} =\mathcal{E}_\mathcal{H}^{(T)} \cup \widehat{\mathcal{E}}_\mathcal{H}^{(T)}$.
This measure provides a more comprehensive evaluation by considering both the presence and frequency of hyperedges.

\smallsection{Comparison with Existing Problems:} 
While there have been efforts to reconstruct hypergraphs from their projected pairwise graphs, most existing studies have overlooked the importance of edge multiplicity and have limited their input data to unweighted projected graphs. 
This approach essentially considers only the existence of pairwise edges, disregarding how frequently the constituent nodes co-appear within hyperedges in the original hypergraph.

To the best of our knowledge, hypergraph reconstruction with edge multiplicity has only been explored by Wang and Kleinberg~\cite{wang2024from}. Specifically, while their main text focuses on supervised hypergraph reconstruction without considering edge multiplicity, their appendix examines an unsupervised approach that incorporates edge multiplicity. However, supervised hypergraph reconstruction with edge multiplicity, which we addressed in this work, has not been explored.

\section{Proposed Method}
\label{sec:method}
We propose \method, an effective method for supervised multiplicity-aware hypergraph reconstruction (i.e., Problem~\ref{problem}).
\method explores cliques in the projected graph as potential hyperedge candidates. 
It employs a classifier trained on multiplicity-aware features to predict the likelihood of each clique being a hyperedge.
For cliques likely to form hyperedges, \method updates the projected graph by removing their constituent edges and incorporates the corresponding sets of nodes as hyperedges in the reconstructed hypergraph.

The core idea of \method is to effectively leverage edge multiplicity information in the projected graph.
Specifically, \method benefits from edge multiplicity in two key ways:

\begin{itemize}[leftmargin=*]
    \item \textbf{Reduced search space:}
    Edge multiplicity significantly reduces the number of potential cases of hyperedge overlaps among the nodes in a given clique.
    This reduction simplifies the hyperedge identification problem and enhances the efficiency of the overall reconstruction process, as discussed in Sect.~\ref{sec:intro}.
    \method not only utilizes edge multiplicity throughout the reconstruction process but also further reduces the search space by leveraging edge multiplicity to identify size-2 hyperedges that are theoretically guaranteed to be true hyperedges in a preprocessing step.
    \item \textbf{Clique feature enhancement:}
    Edge multiplicity represents the number of hyperedges in the original hypergraph that overlap across pairs of nodes.  
    When aggregated across all edges in a clique, this information provides a valuable feature for distinguishing whether the clique corresponds to a true hyperedge in the original hypergraph. \method leverages various features derived from edge multiplicity but a simple classifier trained to predict whether a given clique represents a hyperedge (see Sect.~\ref{sec:proposed:training}).
\end{itemize}

Moreover, \method incorporates a \textbf{bidirectional search} that explores cliques with both high and low predicted likelihoods of being hyperedges.
While it identifies highly probable cliques as hyperedges, it also examines subsets of low-likelihood cliques that might otherwise be overlooked.
This comprehensive approach improves the accuracy of hypergraph reconstruction, as empirically demonstrated in Sect.~\ref{sec:experiments}.

\begin{algorithm}[t]
\small
\DontPrintSemicolon
\caption{Overview of \method \label{algo:main}}
\SetKwInOut{Input}{Input}
\SetKwInOut{Output}{Output}
\SetKwComment{Comment}{$\triangleright$\ }{}

\Input{
    (a) Input projected graph \( \mathcal{G} = (\mathcal{V}, \mathcal{E}_\mathcal{G}, \omega) \) \\
    \hspace{4pt}(b) Multiplicity-aware classifier $\mathcal{M}$ \\
    \hspace{1.4pt} (c) Initial classification threshold $\theta_{\text{init}}$ \\
    \hspace{1.4pt} (d) Negative prediction processing ratio $r$ (\%) \\
    \hspace{1.4pt} (e) Threshold adjust ratio $\alpha$
}
\Output{
    (a) Reconstructed hypergraph $\widehat{\mathcal{H}}$
}

\BlankLine
$\widehat{\mathcal{H}} \leftarrow \emptyset$ \Comment*[r]{Initialize the reconstructed hypergraph}
$\theta \leftarrow \theta_{\text{init}}$\Comment*[r]{\(\theta\) is the classification threshold for $\mathcal{M}$}
$\mathcal{G}', \widehat{\mathcal{H}} \leftarrow \texttt{Filtering}(\mathcal{G}, \widehat{\mathcal{H}})$\label{algo:line:filter}\\ 

\While{$\mathcal{G}'$ has edges}{
    $\mathcal{G}', \widehat{\mathcal{H}} \leftarrow \texttt{BidirectionalSearch}(\mathcal{G}', \mathcal{M}, \theta, r, \widehat{\mathcal{H}})$\label{algo:line:bidirection}\\
    $\theta \leftarrow \max(\theta - \alpha \times \theta_{\text{init}},\; 0)$
    }

\BlankLine
\Return{$\widehat{\mathcal{H}}$}\;

\end{algorithm}

\subsection{\bf Overview of \method (Algorithm~\ref{algo:main})}\label{sec:method:overview}
\method generates a reconstructed hypergraph $\widehat{\mathcal{H}}$, initialized as an empty set, from a given projected graph $\mathcal{G}$.
The reconstruction process comprises two main procedures:

\begin{enumerate}[leftmargin=*]
    \item \textbf{Theoretically-Guaranteed Filtering (Sect.~\ref{sec:proposed:filtering})}: 
    As a preprocessing step, \method identifies edges in $\mathcal{G}$ that are \textit{theoretically guaranteed} to correspond to size-$2$ hyperedges.
    These edges are directly incorporated into the reconstructed hypergraph $\widehat{\mathcal{H}}$ and removed from $\mathcal{G}$, reducing it to an intermediate graph $\mathcal{G}'$.
    \item \textbf{Bidirectional Search (Sect.~\ref{sec:proposed:bidirectional}}):
    \method examines the maximal cliques in $\mathcal{G}'$ as potential hyperedges.
    Each maximal clique is assessed for its likelihood of being a hyperedge using a classifier trained on multiplicity-aware features.
    Based on the predicted likelihood with respect to an adjustable threshold, \method identifies not only (1) cliques with high estimated likelihood but also (2) sub-cliques with high likelihood but hidden within larger low-likelihood cliques, as hyperedges, improving the comprehensiveness of the search.
    This process is iteratively repeated until no edges remain in $\mathcal{G}'$.
\end{enumerate}

Finally, \method completes the process and returns the reconstructed hypergraph $\widehat{\mathcal{H}}$.
Below, we provide a detailed description of each step.

\subsection{\bf Theoretically-Guaranteed Filtering (Algorithm~\ref{algo:filtering})\label{sec:proposed:filtering}}
As a preprocessing step, \method identifies the edges in $\mathcal{G}$ that are \textit{theoretically guaranteed} to correspond to true size-$2$ hyperedges of the original hypergraph $\mathcal{H}$.
These edges are directly incorporated into the reconstructed hypergraph $\mathcal{H}$ and removed from $\mathcal{G}$, resulting in an intermediate graph $\mathcal{G}'$.
This step effectively reduces the search space for subsequent reconstruction steps and allows \method to focus on identifying higher-order (i.e., larger) hyperedges.



\begin{algorithm}[t]
\DontPrintSemicolon
\caption{\texttt{Filtering}\label{algo:filtering}} 
\SetKwInOut{Input}{Input}
\SetKwInOut{Output}{Output}

\Input{
    (a) Input projected graph: \( \mathcal{G} = (\mathcal{V}, \mathcal{E}_\mathcal{G}, \omega) \)\\
    \hspace{3pt}(b) Reconstructed hypergraph \( \widehat{\mathcal{H}} \)\;
}
\Output{
    (a) Intermediate graph \( \mathcal{G}' = (\mathcal{V}, \mathcal{E}_{\mathcal{G}'}, \omega') \)\\
    \hspace{3pt}(b) Updated reconstructed hypergraph \( \widehat{\mathcal{H}} \)\;
}

\BlankLine
$\mathcal{E}_{\mathcal{G}'}\leftarrow \mathcal{E}_{\mathcal{G}}$ and $\omega' \leftarrow \omega$\Comment*[r]{Initialize the intermediate graph}
\For{\textbf{each} edge \( (u, v) \in \mathcal{E}_\mathcal{G} \)}{
    \( \text{MHH}(u, v) \leftarrow \sum_{z \in N(u) \cap N(v)} \min(\omega_{u,z}, \omega_{v,z}) \)\;
    \( r_{u,v} \leftarrow \omega_{u,v} - \text{MHH}(u, v) \)\;

    \If{\( r_{u,v} > 0 \)}{
        \( \widehat{\mathcal{H}} \leftarrow \widehat{\mathcal{H}} \cup \big\{\{u, v\}\big\}^{r_{u,v}} \)\;
        \( \omega_{u,v}^{'} \leftarrow \omega_{u,v} - r_{u,v} \)\;
        \If{\( \omega_{u,v}^{'} = 0 \)}{
            \( \mathcal{E}_{\mathcal{G}'} \leftarrow \mathcal{E}_{\mathcal{G}'} \setminus \{(u, v)\} \)\;
        }
    }
}

\BlankLine
\Return{\( \mathcal{G}' \), \( \widehat{\mathcal{H}} \)}\;

\end{algorithm}

For each edge \( (u, v)\in \mathcal{E}_{\mathcal{G}} \) in $\mathcal{G}$, we define the \textit{maximum number of higher-order hyperedges (\textbf{MHH})} as the maximum possible number of \textit{higher-order hyperedges} (i.e., hyperedges of size $3$ or larger) that involve both \( u \) and \( v \).
Formally, the \textbf{MHH} of nodes $u$ and $v$ is defined as:

\begin{equation}
\text{MHH}(u, v) = \sum_{z \in N(u) \cap N(v)} \min(\omega_{u, z},\ \omega_{v, z})\label{eq:MHH},
\end{equation}
where \( N(u) \) and \( N(v) \) represent the set of neighbors of $u$ and $v$, respectively.
From the definition of MHH, we derive Lemma~\ref{lemma:MHH_1} and Lemma~\ref{lemma:MHH_2}, which provide theoretical guarantees for identifying size-$2$ hyperedges.

\begin{lemma}[Upper Bound of Higher-Order Hyperedges]\label{lemma:MHH_1}
The $\text{MHH}(u, v)$ is an upper bound on the total number of higher-order hyperedges that include both \( u \) and \( v \).
\end{lemma}

From Lemma~\ref{lemma:MHH_1}, we define the residual edge multiplicity $r_{u,v}$ between nodes $u$ and $v$ as follows:

\[
r_{u,v} = \omega_{u,v} - \text{MHH}(u, v),
\]

\noindent
which directly leads to Lemma~\ref{lemma:MHH_2}.

\begin{lemma}[Lower Bound on Size-2 Hyperedges]\label{lemma:MHH_2}
The residual edge multiplicity \( r_{u,v} \) is a lower bound on the number of true hyperedges in \( \mathcal{H} \) that exclusively consist of $u$ and $v$.
\end{lemma}

Therefore, if \( r_{u,v} > 0 \), we can theoretically and confidently identify \( \{u, v\} \) as a true size-$2$ hyperedge.
Such edges are removed from $\mathcal{G}$ and incorporated into the reconstructed hypergraph $\widehat{\mathcal{H}}$.
The proofs of Lemma~\ref{lemma:MHH_1} and Lemma~\ref{lemma:MHH_2} are provided in the online appendix\cite{supple}.

\begin{algorithm}[t]
\DontPrintSemicolon
\caption{\texttt{BidirectionalSearch} \label{algo:bidirection}}
\SetKwInOut{Input}{Input}
\SetKwInOut{Output}{Output}
\SetKwComment{Comment}{$\triangleright$\ }{}

\Input{
    (a) Intermediate graph \( \mathcal{G}' = (\mathcal{V}, \mathcal{E}_{\mathcal{G}'}, \omega') \)\\
    \hspace{3pt}(b) Multiplicity-aware classifier \( \mathcal{M} \)\\
    \hspace{3pt}(c) Classification threshold \( \theta \)\\
    \hspace{3pt}(d) Reconstructed hypergraph \( \widehat{\mathcal{H}} \)\\
    \hspace{3pt}(e) Negative prediction processing ratio \( r \) (\%) \\
}
\Output{
    (a) Updated intermediate graph \( \mathcal{G}' = (\mathcal{V}, \mathcal{E}_{\mathcal{G}'}, \omega') \)\\
    \hspace{3pt}(b) Updated reconstructed hypergraph \( \widehat{\mathcal{H}} \)
}

\BlankLine
\( \mathcal{Q} \leftarrow \text{All maximal cliques (sets of nodes) in } \mathcal{G}' \)\;
\( \mathcal{Q}_{\text{pos}} \leftarrow \{ Q \in \mathcal{Q} : \mathcal{M}(Q) > \theta \} \)\;
\( \mathcal{Q}_{\text{neg}} \leftarrow \{ Q \in \mathcal{Q} \setminus \mathcal{Q}_{\text{pos}} : \text{Lowest } r\% \text{ w.r.t. } \mathcal{M}(Q) \} \)\;

\Comment{Phase 1: Most Promising Cliques}
Sort \( \mathcal{Q}_{\text{pos}} \) in descending order of \( \mathcal{M}(Q) \)\;

\For{\textbf{each} clique \( Q \in \mathcal{Q}_{\text{pos}} \)}{
     \( \mathcal{E}_Q \leftarrow \{(u, v) : u, v \in Q,\ u \neq v \} \)\;
    \If{\( \mathcal{E}_Q \subseteq \mathcal{E}_{\mathcal{G}'} \)}{
        \( \widehat{\mathcal{H}} \leftarrow \widehat{\mathcal{H}} \cup \{Q\} \)\;
        \( \omega_{u, v}^{'} \leftarrow \omega_{u, v}^{'} - 1 \text{ for all } (u, v) \in \mathcal{E}_Q  \)\;
        \( \mathcal{E}_{\mathcal{G}'} \leftarrow \mathcal{E}_{\mathcal{G}'} \setminus \{(u, v) \in \mathcal{E}_Q : \omega_{u,v}^{'} = 0 \} \)\;
    }
}

\BlankLine
\Comment{Phase 2: Least Promising Cliques}

\( \overline{\mathcal{Q}}_{\text{sub}} \leftarrow \bigcup_{Q \in \mathcal{Q}_{\text{neg}}} \bigcup_{k=2}^{|Q|-1} \text{RandomSample}(\binom{Q}{k}) \)\;
\( \mathcal{Q}_{\text{sub}} \leftarrow \{ Q \in \overline{\mathcal{Q}}_{\text{sub}} : \mathcal{M}(Q) > \theta \} \)\;
Sort \( \mathcal{Q}_{\text{sub}} \) in descending order of \( \mathcal{M}(Q) \)\;

\For{\textbf{each} subset \( Q_{\text{sub}} \in \mathcal{Q}_{\text{sub}} \)}{
    \( \mathcal{E}_{Q_{\text{sub}}} \leftarrow \{(u, v) : u, v \in Q_{\text{sub}},\ u \neq v \} \)\;
    \If{\( \mathcal{E}_{Q_{\text{sub}}} \subseteq \mathcal{E}_{\mathcal{G}'} \)}{
        \( \widehat{\mathcal{H}} \leftarrow \widehat{\mathcal{H}} \cup \{Q_{\text{sub}}\} \)\;
        \( \omega_{u,v}^{'} \leftarrow \omega_{u,v}^{'} - 1 \text{ for all } (u, v) \in \mathcal{E}_{Q_{\text{sub}}} \)\;
        \( \mathcal{E}_{\mathcal{G}'} \leftarrow \mathcal{E}_{\mathcal{G}'} \setminus \{(u, v) \in \mathcal{E}_{Q_{\text{sub}}} : \omega_{u,v}^{'} = 0 \} \)\;
    }
}

\BlankLine
\Return{\( \mathcal{G}' \), \( \widehat{\mathcal{H}} \)}\;

\end{algorithm}

\subsection{\bf Bidirectional Search (Algorithm~\ref{algo:bidirection})}\label{sec:proposed:bidirectional}
\method examines the cliques in the intermediate graph \( \mathcal{G}' \) to determine whether they qualify as hyperedges.
It employs a bidirectional search, which not only evaluates cliques with high predicted likelihoods as hyperedges but also investigates sub-cliques within low-likelihood cliques to uncover potential hyperedges that might otherwise be overlooked.

Specifically, \method identifies all maximal cliques \( \mathcal{Q} \) in \( \mathcal{G}' \), and for each maximal clique $Q\in \mathcal{Q}$, it uses a classifier \( \mathcal{M} \), trained on multiplicity-aware features (as detailed in Sect.~\ref{sec:proposed:training}), to predict the score $\mathcal{M}(Q)$ of the clique being a hyperedge.
Based on the predicted scores relative to the threshold $\theta$, which is adaptively adjusted throughout the process, the maximal cliques are categorized into two subsets: (1) the set of \textbf{\textit{most promising cliques} $\mathcal{Q}_{\text{pos}}$}, consisting of cliques with prediction scores above the threshold $\theta$ (i.e., $\mathcal{M}(Q) >\theta \;\forall Q\in \mathcal{Q}_{\text{pos}}$); and (2) the set of \textbf{\textit{least promising cliques} $\mathcal{Q}_{\text{neg}}$}, consisting of the cliques with the lowest $r$\% prediction scores among $\mathcal{Q} \setminus \mathcal{Q}_{\text{pos}}$.
The bidirectional search in Algorithm~\ref{algo:bidirection} applies distinct approaches to each subset of cliques, as discussed below.



\smallsection{Most Promising Cliques:}
The cliques with high prediction scores are \textit{sequentially} added to the reconstructed hypergraph \( \widehat{\mathcal{H}} \) as hyperedges. 
Simultaneously, their constituent edges are removed from \( \mathcal{G}' \).
It is important to note that if a prioritized clique (e.g., $\{5,6,7\}$ in Fig.~\ref{fig:model}) is removed from $\mathcal{G}'$, a subsequent clique (e.g., $\{2,3,5,6\}$ in Fig.~\ref{fig:model}) may no longer exist in the updated intermediate graph $\mathcal{G}'$ because its overlapping edges with the former clique have already been removed.

\smallsection{Least Promising Cliques:}
For cliques with low prediction scores, which are predicted as unlikely to be hyperedges, it is still possible that they contain sub-cliques that are true hyperedges in the original hypergraph $\mathcal{H}$.
To identify such sub-cliques, for each clique $Q \in \mathcal{Q}_{\text{neg}}$, and for each sub-clique size $k\in \{2, ..., |Q|-1\}$, \method randomly samples sub-cliques from the set of all possible sub-cliques (i.e., $\binom{Q}{k}$). 
The resulting set of sampled sub-cliques, denoted as $\overline{\mathcal{Q}}_\text{sub}$, is formally defined as:
\begin{equation*}
    \overline{\mathcal{Q}}_\text{sub} = \bigcup_{Q\in \mathcal{Q}_\text{neg}} \bigcup_{k=2}^{|Q|-1} \text{RandomSample}(\binom{Q}{k}),
\end{equation*}
where $\text{RandomSample}(\cdot)$ randomly samples an element from the input set. 
From the set $\overline{\mathcal{Q}}_\text{sub}$, which contains $\sum_{Q\in \mathcal{Q}_\text{neg}}(|Q|-2)$ sub-cliques, those with high prediction scores are filtered to produce $\mathcal{Q}_\text{sub}$, defined as:
\begin{equation*}
    \mathcal{Q}_\text{sub} = \{ Q\in \overline{\mathcal{Q}}_\text{sub} : \mathcal{M}(Q) > \theta \}.
\end{equation*}
The sub-cliques in $\mathcal{Q}_\text{sub}$, which are estimated to be true hyperedges, are added to the reconstructed hypergraph $\widehat{\mathcal{H}}$.
Simultaneously, the intermediate graph $\mathcal{G}'$ is updated by removing the constituent edges of the removed sub-cliques. 
In addition, identifying sub-cliques within the least promising cliques ensures that these cliques do not reappear in subsequent iterations and thus enhances the efficiency of \method.

\smallsection{Adaptive Threshold Adjustment:} 
The threshold \( \theta \), which is used to estimate the likelihood of the clique $Q$ being a hyperedge based on the predicted score $\mathcal{M}(Q)$, is adaptively decreased over iterations using a threshold adjustment ratio $\alpha$. 
Specifically, after each iteration, $\theta$ is updated as follows:
\begin{equation*}
    \theta \leftarrow \max(\theta - \alpha \times \theta_{\text{init}},\ 0),
\end{equation*}
where $\theta_\text{init}$ is the initial value of the threshold. 
This adaptive adjustment allows more candidates to be considered as potential hyperedges in subsequent iterations. By gradually lowering the threshold, the recall of the reconstructed hypergraph is enhanced, as it captures cliques that may have lower prediction scores but are still likely to be hyperedges.

\subsection{\bf Multiplicity-Aware Classifier}\label{sec:proposed:training}
Lastly, we discuss how the classifier $\mathcal{M}$, which generates the prediction scores $\mathcal{M}(Q)$ for a given clique $Q$, is trained.
Specifically, $\mathcal{M}$ is trained to determine whether a clique $Q$ in the source projected graph $\mathcal{G}^{(S)}$ corresponds to a hyperedge in the source hypergraph $\mathcal{H}^{(S)}$.
The classifier leverages a carefully designed feature representation for each clique, which comprehensively captures edge multiplicity information.


\smallsection{Multiplicity-Aware Clique Features:}
For a clique $Q$ identified in the projected graph, we generate its feature representation.
Specifically, we integrate three levels of features: node-level, edge-level, and clique-level features, which account for edge multiplicity.

\begin{itemize}[leftmargin=*]
    \item \textbf{Node-level features:} 
    For each node $u\in Q$ in the clique $Q$, we use its weighted degree (i.e., $\sum_{v\in N(u)}\omega_{u,v}$).
    \item \textbf{Edge-level features:}
    For each edge $(u,v)\in \mathcal{E}_Q$ within the clique $Q$, we include (1) its multiplicity (i.e., $\omega_{u,v}$), (2) the maximum number of higher-order hyperedges (i.e., $\text{MHH}(u,v)$; see Eq.~\eqref{eq:MHH}), and (3) the \textit{maximum portion of higher-order hyperedges} (i.e., $\text{MHH}(u,v) / \omega_{u,v}$), which quantifies the maximum proportion of the higher-order hyperedges that contribute to the edge multiplicity $\omega_{u,v}$.
    \item \textbf{Clique-level features:}
    For the clique $Q$, we include (1) its clique size (i.e., $|Q|$), (2) the \textit{clique cut ratio}, which measures the proportion of edge multiplicity within the clique relative to the total edge multiplicity connected to nodes in the clique, and (3) a binary indicator specifying whether the clique is maximal in the projected graph $\mathcal{G}$ (i.e., $1$ if $Q$ is maximal in $\mathcal{G}$, and $0$ otherwise).
\end{itemize}

To summarize node-level and edge-level features into clique-level features, we aggregate the respective node or edge features for each clique. 
Specifically, for each set of node or edge features, we calculate the sum, average, minimum, maximum, and standard deviation, resulting in a $5$-dimensional vector. 
These aggregated vectors are then concatenated with the predefined clique-level features to form the final feature representation for the clique.
In Sect.~\ref{sec:exp:multiplicity}, we empirically demonstrate that these features derived from edge multiplicity are more effective and informative than other potentially feasible clique feature representations.

The classifier $\mathcal{M}$ is implemented as a simple MLP.
For further details about $\mathcal{M}$, including its negative sampling strategy, refer to the online appendix \cite{supple}.

\subsection{\bf Complexity Analysis}\label{sec:complexity}
We analyze the time and space complexity of \method. 
All proofs can be found in the online appendix\cite{supple}. 

\smallsection{Time Complexity:} 
We first analyze the time complexity of \method when applied to the graph $\mathcal{G}=(\mathcal{V},\mathcal{E}_\mathcal{G},\omega)$.
Specifically, we analyze the complexities of the \texttt{Filtering} step (Algorithm~\ref{algo:filtering}) and \texttt{BidirectionalSearch} step (Algorithm~\ref{algo:bidirection}). For the complexity of the negative sampling step for training the classifier $\mathcal{M}$, refer to \cite{supple}.
The empirical runtimes of \method are presented in Sect.~\ref{sec:exp:runningtime}.

\begin{lemma}[Time Complexity of Algorithm~\ref{algo:filtering}]\label{lemma:time:filtering}
The time complexity of the \texttt{Filtering} step is $O(|\mathcal{E}_\mathcal{G}| \cdot d_{\max})$, where $d_{\max}$ is the maximum degree of \(\mathcal{G}\), i.e., $d_{\max}=\max_{u\in \mathcal{V}} |N(u)|$.
\end{lemma}

Assuming that $d_{\max}$ is a constant, the \texttt{Filtering} step exhibits a practical time complexity of $O(|\mathcal{E}_\mathcal{G}|)$.

\begin{lemma}[Time Complexity of Algorithm~\ref{algo:bidirection}]\label{lemma:time:search}
The time complexity of the \texttt{BidirectionalSearch} step is $O\left( N \log |\mathcal{C}| \right)$, where $N = \sum_{C\in \mathcal{C}} m_C$.
Here, $\mathcal{C}$ is the set of all cliques in $\mathcal{G}$, and $m_C=\min_{(u,v)\in C}\omega_{u,v}$ is the minimum edge multiplicity among the edges in clique $C$.
\end{lemma}

\smallsection{Space Complexity:} 
\method requires  $O(|\mathcal{V}|+|\mathcal{E}_\mathcal{G}|)$ space for the input projected graph,  and it requires extra space proportional to the sum of sizes (i.e., node counts) of the cliques processed in each iteration of the \texttt{BidrectionalSearch} step, which is a subset of all cliques in $\mathcal{G}$.

\section{Experiments}
\label{sec:experiments}
We perform experiments to answer the following questions:
\begin{enumerate}[leftmargin=*]
    \item[Q1.] \textbf{Accuracy.}
    How accurately does \method reconstruct the projected graph into hypergraphs? 
    How effectively do the reconstructed hypergraphs preserve the structural properties of the original hypergraph?
    \item[Q2.] \textbf{Transferability.}
    Can \method, trained on one dataset, reconstruct other hypergraphs from the same data domain?
    \item[Q3.] \textbf{Applicability.}
    How much does hypergraph reconstruction by \method impact downstream task performance?
    \item[Q4.] \textbf{Effectiveness.}
    Is every component of \method effective?
    Which clique features are most important? 
    Is \method robust to changes in its hyperparameters?
    \item[Q5.] \textbf{Scalability.}
    How fast is \method, compared to competitors?
    How does its running time scale with the graph size?
\end{enumerate}


\subsection{\bf Experimental Settings}
\label{sec:experiments:settings}


\begin{table}[t]
\centering
\caption{{\small{\underline{\smash{Summary of the datasets.}} 
We report the number of nodes ($|\mathcal{V}|$), the number of hyperedges ($|\mathcal{E}_\mathcal{H}|$), and the average hyperedge multiplicity ($\text{Avg. }M_\mathcal{H}$) in the original hypergraph $\mathcal{H}$, as well as the number of edges (|$\mathcal{E}_{\mathcal{G}}$|) and the average edge multiplicity ($\text{Avg. }\omega$) in the corresponding projected graph $\mathcal{G}$.
}}}
\label{tab:DatasetTable}
\setlength\tabcolsep{5.6pt}
\scalebox{0.945}{
\begin{tabular}{l|c|cc|cc}
\toprule
& \textbf{Nodes} & \multicolumn{2}{c|}{\textbf{Hyperedges in $\mathcal{H}$}} & \multicolumn{2}{c}{\textbf{Edges in $\mathcal{G}$}}\\
\textbf{Dataset} & \textbf{$|\mathcal{V}|$} & \textbf{$|\mathcal{E}_\mathcal{H}|$} & \textbf{$\text{Avg. } {M_\mathcal{H}}$} & \textbf{$|\mathcal{E}_\mathcal{G}|$} & \textbf{$\text{Avg. } \omega$}\\
\midrule
Enron~\cite{benson2018simplicial}        & 141  & 889  & 5.85 & 5,205 & 9.18\\
P. School~\cite{benson2018simplicial}    & 238   & 7,975   & 6.90  & 55,043 & 11.98 \\
H. School~\cite{benson2018simplicial}    & 318   & 4,254   & 17.01 & 72,369 & 22.24\\
Crime~\cite{young2021hypergraph}         & 308  & 105  & 1.01  & 106   & 1.03\\
Hosts~\cite{young2021hypergraph}        & 449  & 159   & 1.06 & 168   &  1.24\\
Directors~\cite{young2021hypergraph}     & 513   & 101    &  1.01 & 102  & 1.02\\
Foursquare~\cite{young2021hypergraph}    & 2,254  & 873   & 1.00  & 873  & 1.02\\
DBLP~\cite{benson2018simplicial}         & 389,330 & 213,328 & 1.10  & 235,498 & 1.28\\
Eu~\cite{benson2018simplicial}           & 891  & 6,805   & 1.26 & 8,581 & 4.62\\
MAG-TopCS~\cite{amburg2020clustering}    & 48,742 & 25,945  & 1.00  & 25,945 & 1.14\\
\bottomrule
\end{tabular}%
}
\end{table}

\begin{table*}[t]
\vspace{-2mm}
\centering 
\caption{\small{\underline{\smash{Reconstruction accuracy in the \textbf{multiplicity-reduced setting}.}} 
\method demonstrates the highest Jaccard similarity (scaled by a factor of 100) compared to the baseline methods.
The best performance is shown in \textbf{bold}, and the second best is \underline{\smash{underlined}}. 
"OOT" indicates "out of time" (exceeding 24 hours), and "OOM" represents "out of memory" (system limit: 384 GB).}}
\setlength\tabcolsep{4.0pt}
\scalebox{0.84}{
\begin{tabular}{c| c c c c c c c c c c}
\toprule
\textbf{Method} & \textbf{Enron} & \textbf{P.School} & \textbf{H.School} & \textbf{Crime} & \textbf{Hosts} & \textbf{Directors} & \textbf{Foursquare} & \textbf{DBLP} & \textbf{Eu} & \textbf{MAG-TopCS} \\
\midrule
\cfinder                  & 0.00 $\pm$ 0.00 & 0.00 $\pm$ 0.00& 0.00 $\pm$ 0.00& 24.04 $\pm$ 0.00& 2.50 $\pm$ 0.00& 41.18 $\pm$ 0.00& 7.81 $\pm$ 0.00& 21.50 $\pm$ 0.00& 0.01 $\pm$ 0.00& 25.34 $\pm$ 0.00\\
\demon                    & 2.43 $\pm$ 0.00 & 0.09 $\pm$ 0.00 & 2.97 $\pm$ 0.00 & 76.27 $\pm$ 0.48 & 9.21 $\pm$ 0.62 & 90.14 $\pm$ 1.39 & 17.01 $\pm$ 0.35 & 49.05 $\pm$ 0.02 & 0.01 $\pm$ 0.00 & 24.75 $\pm$ 0.05 \\
\maxclique                & 4.31 $\pm$ 0.00& 0.09 $\pm$ 0.00& 2.38 $\pm$ 0.00& 92.82 $\pm$ 0.00& 23.19 $\pm$ 0.00& \textbf{100.00 $\pm$ 0.00}& 9.55 $\pm$ 0.00& 84.51 $\pm$ 0.00& 0.98 $\pm$ 0.00& 82.12 $\pm$ 0.00\\
\ecc                      & 6.84 $\pm$ 0.00& 1.95 $\pm$ 0.00& 6.89 $\pm$ 0.00& 93.24 $\pm$ 0.00& 54.19 $\pm$ 0.00& \textbf{100.00 $\pm$ 0.00}& 93.69 $\pm$ 0.00& 83.87 $\pm$ 0.00& 7.11 $\pm$ 0.00& 84.27 $\pm$ 0.00\\
\bayesian                 & 4.77 $\pm$ 0.26 & 0.18 $\pm$ 0.02 & 3.57 $\pm$ 0.05 & 93.15 $\pm$ 0.20 & 53.10 $\pm$ 0.32 & \textbf{100.00 $\pm$ 0.00} & 80.65 $\pm$ 1.06 & 86.06 $\pm$ 0.01 & 4.76 $\pm$ 0.03 & 87.26 $\pm$ 0.02 \\
\wangunsup                & 13.74 $\pm$ 0.00& 8.34 $\pm$ 0.00& 17.00 $\pm$ 0.00& 94.86 $\pm$ 0.00& 50.38 $\pm$ 0.00& \textbf{100.00 $\pm$ 0.00}& 92.27 $\pm$ 0.00& OOT & 5.19 $\pm$ 0.00& 94.31 $\pm$ 0.00\\
\wang-Motif               & 14.14 $\pm$ 3.27 & OOT & 54.21 $\pm$ 0.31 & 92.82 $\pm$ 0.00 & 49.69 $\pm$ 0.59 & \textbf{100.00 $\pm$ 0.00} & 70.92 $\pm$ 6.65 & 85.99 $\pm$ 0.04 & OOM & 86.04 $\pm$ 0.11\\
\wang-Count               & 14.36 $\pm$ 0.50 & 42.87 $\pm$ 1.66 & 54.19 $\pm$ 0.22 & 92.82 $\pm$ 0.00 & 56.64 $\pm$ 0.55 & \textbf{100.00 $\pm$ 0.00} & 85.96 $\pm$ 0.06 & 86.07 $\pm$ 0.00 & 11.14 $\pm$ 0.27  & 87.14 $\pm$ 0.09 \\
\midrule
\method-M          & 19.48 $\pm$ 0.45 & 47.04 $\pm$ 0.56 & 55.69 $\pm$ 0.18 & 93.65 $\pm$ 0.30 & 56.17 $\pm$ 2.64 & \textbf{100.00 $\pm$ 0.00} & 94.03 $\pm$ 3.04 & 95.86 $\pm$ 0.44 & 11.85 $\pm$ 0.05 & 96.57 $\pm$ 9.24 \\
\method-F          & \underline{22.95 $\pm$ 0.92} & \underline{47.26 $\pm$ 0.64} & \underline{57.66 $\pm$ 0.40} & 96.33 $\pm$ 7.35 & 57.47 $\pm$ 3.46 & \textbf{100.00 $\pm$ 0.00} & 97.27 $\pm$1.80 & 97.82 $\pm$ 0.11 & \underline{12.48 $\pm$ 0.41} & 97.01 $\pm$ 0.12 \\
\method-B            & 21.23 $\pm$ 1.79 & 9.59 $\pm$ 0.38 & 18.04 $\pm$ 0.32 & \textbf{100.00 $\pm$ 0.00} & \underline{60.08 $\pm$ 5.34} & \textbf{100.00 $\pm$ 0.00} & \textbf{100.00 $\pm$ 0.00} & \textbf{98.25 $\pm$ 0.02} & 11.87 $\pm$ 0.13 & \textbf{98.20 $\pm$ 0.09} \\
\textbf{\method}            & \textbf{25.06 $\pm$ 0.60} & \textbf{47.48 $\pm$ 0.65} & \textbf{57.75 $\pm$ 0.29} & \textbf{100.00 $\pm$ 0.00 }&\textbf{ 61.52 $\pm$ 2.63} & \textbf{100.00 $\pm$ 0.00} & \underline{99.19 $\pm$ 0.34} & \underline{98.13 $\pm$ 0.06} & \textbf{13.98 $\pm$ 0.23}  & \underline{98.04 $\pm$ 0.12} \\
\bottomrule
\end{tabular}%
}
\label{tab:jaccard_comparison}
\end{table*}

\begin{table*}[t]
\centering

\caption{\small{\underline{\smash{Reconstruction accuracy in the \textbf{multiplicity-preserved setting}.}} 
\method demonstrates the highest multi-Jaccard similarity (scaled by a factor of 100) compared to the baseline methods.
The best performance is shown in \textbf{bold}, and the second best is \underline{\smash{underlined}}. 
"OOT" indicates "out of time" (exceeding 24 hours).}}
\setlength\tabcolsep{4.0pt}
\scalebox{0.84}{
\begin{tabular}{c| c c c c c c c c c c}
\toprule
\textbf{Method} & \textbf{Enron} & \textbf{P.School} & \textbf{H.School} & \textbf{Crime} & \textbf{Hosts} & \textbf{Directors} & \textbf{Foursquare} & \textbf{DBLP} & \textbf{Eu} & \textbf{MAG-TopCS} \\
\midrule
\bayesian                 & 3.51 $\pm$ 0.16 & 0.11 $\pm$ 0.02 & 2.38 $\pm$ 0.11 & 95.06 $\pm$ 0.40 & 56.25 $\pm$ 1.29 & \textbf{100.00 $\pm$ 0.00 }& 80.18 $\pm$ 0.76 & 85.10 $\pm$ 0.01 & 4.92 $\pm$ 0.05 & 87.36 $\pm$ 0.03 \\
\wangunsup   & $43.45$ $\pm$ 0.00 & $25.30$ $\pm$ 0.00 & $56.76$ $\pm$ 0.00 & \textbf{100.00 $\pm$ 0.00} & $12.39$ $\pm$ 0.00 & \textbf{100.00 $\pm$ 0.00} & $97.97$ $\pm$ 0.00 & OOT & 5.62 $\pm$ 0.00 & 93.51 $\pm$ 0.00 \\
\midrule
\method-M          & \underline{48.89 $\pm$ 0.40} & 51.49 $\pm$ 0.47 & 68.66 $\pm$ 0.21 & 91.84 $\pm$ 2.11 & 50.32 $\pm$ 12.86 & \textbf{100.00 $\pm$ 0.00} & 84.49 $\pm$ 1.74 & 92.84 $\pm$ 1.78 & \underline{10.63 $\pm$ 0.75}  & 95.73 $\pm$ 0.12 \\
\method-F          & 48.37 $\pm$ 0.72  & \underline{53.26 $\pm$ 0.80} & \underline{69.29 $\pm$ 0.25} & 96.33 $\pm$ 0.00 & 47.16 $\pm$ 16.38 & 93.91 $\pm$ 6.31 & 96.24 $\pm$ 4.51 & 97.08 $\pm$ 0.03 & 10.18 $\pm$ 0.54 & 96.18 $\pm$ 0.10 \\
\method-B            & 47.61 $\pm$ 0.19 & 27.81 $\pm$ 0.10 & 58.14 $\pm$ 0.05 & \textbf{100.00 $\pm$ 0.00} & \underline{54.63 $\pm$ 19.68} & \textbf{100.00 $\pm$ 0.00} & \textbf{99.01 $\pm$ 0.99} & \textbf{97.49 $\pm$ 0.07} & 9.88 $\pm$ 0.41 & \textbf{97.76 $\pm$ 0.06} \\
\textbf{\method}         & \textbf{52.26 $\pm$ 1.54} & \textbf{53.48 $\pm$ 1.42} & \textbf{69.85 $\pm$ 0.21} & \textbf{100.00 $\pm$ 0.00} & \textbf{58.41 $\pm$ 15.22} & \textbf{100.00 $\pm$ 0.00} & \underline{$98.88 \pm 0.46$} & \underline{$97.48 \pm 0.05$} & \textbf{11.55 $\pm$ 0.52} & \underline{97.62 $\pm$ 0.03} \\
\bottomrule
\end{tabular}%
}
\label{tab:mjaccard_comparison}
\end{table*}

\smallsection{Datasets:}
We use $10$ datasets from various domains, summarized in Table~\ref{tab:DatasetTable}.
For each hypergraph $\mathcal{H}=(\mathcal{V}, \mathcal{E}_{\mathcal{H}}^{*})$, we split its hyperedges into halves, $\mathcal{E}_{\mathcal{H}}^{*(S)}$ and $\mathcal{E}_{\mathcal{H}}^{*(T)}$, to construct the source hypergraph $\mathcal{H}^{(S)}=(\mathcal{V}^{(S)}, \mathcal{E}_{\mathcal{H}}^{*(S)})$ and the target hypergraph $\mathcal{H}^{(T)}=(\mathcal{V}^{(T)}, \mathcal{E}_{\mathcal{H}}^{*(T)})$.
Specifically, for datasets where timestamps of each hyperedge are available, the hyperedges are split into halves based on their timestamps; otherwise, they are randomly split.
The source hypergraph $\mathcal{H}^{(S)}$ is projected into $\mathcal{G}^{(S)}$ which is used to train \method. 


To align with the experimental setups of previous studies~\cite{young2021hypergraph, wang2024from}, we use \textit{multiplicity-reduced} hypergraphs, where the multiplicities of all hyperedges are reduced to $1$, i.e., $M_\mathcal{H}(e)=1 \;\forall e\in \mathcal{E}_\mathcal{H}$, unless stated otherwise.
Note that this does not reduce the edge multiplicities in the projected graph to $1$.
In addition, we shall also evaluate \method on \textit{multiplicity-preserved} hypergraphs, where hyperedge multiplicities are not reduced.

\smallsection{Baselines:}
We compare \method with seven baseline methods across three categories:
(1) overlapping community detection-based methods (\demon~\cite{coscia2012demon} and \cfinder~\cite{palla2005uncovering}), 
(2) clique decomposition-based methods (\ecc\cite{conte2016clique} and \maxclique\cite{bron1973algorithm}),
(3) hypergraph reconstruction methods (\wang-Count~\cite{wang2024from}, \wang-Motif~\cite{wang2024from}, \wangunsup~\cite{wang2024from}, and \bayesian~\cite{young2021hypergraph}).
Refer to\cite{supple} for detailed descriptions of each method.


\smallsection{Implementation:}
We implemented \method, \demon, \cfinder, and \wangunsup in Python.
For the hyperparameters required in each method, we set: \method with $\alpha=1/20$ unless otherwise specified, 
\demon with a minimum community size of $2$ and $\epsilon=1$, and
\cfinder with the optimal $k$ selected within the $[0.1, 0.5]$ quantile range of the hyperedge sizes.
For \ecc, \maxclique, and \bayesian we used their official libraries in graph-tools with default hyperparameters.
For \wang, we used hyperparameters recommended by the authors in the original paper.
To ensure fairness, the same maximal clique detection algorithm was used across all methods.

\smallsection{Machines}: All experiments are conducted on a server with dual 2.4GHz Intel Xeon Silver 4210R processors, 384GB memory, and 4 NVIDIA RTX A6000 GPUs.

\subsection{\bf Q1. Accuracy}
Firstly, and most importantly, we evaluate the reconstruction quality of \method against its baselines.

\smallsection{Reconstruction Accuracy:}
In Table~\ref{tab:jaccard_comparison}, we measure the Jaccard similarity between the sets of hyperedges in the target hypergraph and those in the reconstructed hypergraph for the multiplicity-reduced setting.
In Table~\ref{tab:mjaccard_comparison}, we measure the multi-Jaccard similarity for reconstructing multiplicity-preserved hypergraphs, where hyperedge multiplicities are considered.
Notably, only some baselines are applicable for reconstructing multiplicity-preserved hypergraphs.
In both settings, \method consistently and significantly outperforms its competitors across all datasets. 
For example, in the multiplicity-reduced setting, on the Enron dataset, \method yields a reconstructed hypergraph with a Jaccard similarity that is 74.51\% higher than that of \wang-Count. Similarly, in the multiplicity-preserved setting, on the P.School dataset, \method attains a multi-Jaccard similarity that is 111.38\% higher than that of \wangunsup.

\smallsection{Case Studies:}\label{exp:case}
In Fig.~\ref{fig:case_study}, we present the results on the DBLP dataset, focusing on a sub-hypergraph induced by Jure Leskovec and his ten randomly selected co-authors.
Specifically, the co-authorship hypergraph in 2015 is used for training to reconstruct that in 2017.
As shown in the figure, \method exactly restores the sub-hypergraph from its projected graph (Jaccard and multi-Jaccard similarity $= 1.000$), while \wang-Count \cite{wang2024from}  recovers only a subset of ground-truth hyperedges along with some false positives (Jaccard similarity $= 0.333$).
Refer to the online appendix\cite{supple} for
more case studies on the Host–Virus and Crime datasets.

\begin{table*}[h]
\vspace{-2mm}
\centering
\caption{\small{\underline{\smash{Preservation of structural properties.}} 
\method reconstructs hypergraphs with the lowest preservation error (lower is better) across 12 scalar and distributional structural properties.
This indicates \method's strong capability to preserve the structural properties of the original hypergraph.
The best performance is shown in \textbf{bold}, and the second best is \underline{\smash{underlined}}.\label{tab:reconstruction_analysis}}}
\setlength\tabcolsep{6.9pt}
\scalebox{0.96}{
\begin{tabular}{cl|ccccc}
    \toprule
    \multicolumn{2}{c|}{\textbf{Structural Properties}} & \textbf{\bayesian} & \textbf{\wang-Count} & \textbf{\wang-Motif} & \textbf{\wangunsup} & \textbf{\method} \\
    \midrule
    \multirow{7}{*}{\textbf{Scalar Properties}} & Number of Nodes & \textbf{0.000 ± 0.000} & 0.013 ± 0.033 & 0.007 ± 0.017 & \textbf{0.000 ± 0.000} & \textbf{0.000 ± 0.000} \\
    & Number of Hyperedges          & 0.279 ± 0.339 & 0.180 ± 0.235 & \underline{\smash{0.166 ± 0.211}} & 0.167 ± 0.198 & \textbf{0.087 ± 0.134} \\
    & Average Node Degree            & 0.178 ± 0.247 & 0.151 ± 0.236 & 0.161 ± 0.227 & \underline{\smash{0.095 ± 0.149}} & \textbf{0.044 ± 0.070} \\
    & Average Hyperedge Size        & 0.187 ± 0.223 & \underline{\smash{0.085 ± 0.086}} & 0.087 ± 0.091 & 0.089 ± 0.093 & \textbf{0.060 ± 0.079}\\
    & Simplicial Closure Ratio   & 0.151 ± 0.314 & 0.178 ± 0.372 & 0.179 ± 0.368 & \underline{\smash{0.139 ± 0.236}} & \textbf{0.107 ± 0.186} \\
    & Hypergraph Density                    & 0.279 ± 0.339 & 0.175 ± 0.224 & \underline{\smash{0.163 ± 0.205}} & 0.167 ± 0.198 & \textbf{0.087 ± 0.134} \\
    & Hypergraph Overlapness                & 0.178 ± 0.247 & 0.151 ± 0.236 & 0.161 ± 0.227 & \underline{\smash{0.095 ± 0.149}} & \textbf{0.044 ± 0.070} \\
    \midrule
    \multirow{5}{*}{\textbf{Distributional Properties}} & Node Degree        & 0.154 ± 0.235 & 0.118 ± 0.196 & 0.128 ± 0.192 & \underline{\smash{0.081 ± 0.155}} & \textbf{0.033 ± 0.055} \\
    & Node-Pair Degree   & 0.059 ± 0.073 & 0.087 ± 0.133 & 0.101 ± 0.128 & \textbf{0.011 ± 0.020} & \underline{\smash{0.025 ± 0.048}} \\
    & Node-Triple Degree & 0.052 ± 0.071 & 0.155 ± 0.373 & 0.041 ± 0.049 & \underline{\smash{0.016 ± 0.027}} & \textbf{0.015 ± 0.030} \\
    & Hyperedge Homogeneity      & 0.151 ± 0.118 & 0.213 ± 0.229 & 0.241 ± 0.230 & \textbf{0.055 ± 0.068} & \underline{\smash{0.073 ± 0.104}} \\
    & Singular Values & 0.035 ± 0.036 & 0.098 ± 0.124 & 0.102 ± 0.118 & \textbf{0.034 ± 0.038} & \textbf{0.034 ± 0.038} \\
    \midrule
    & \textbf{Average (Overall)}           & 0.142 ± 0.090 & 0.134 ± 0.055 & 0.128 ± 0.064 & \underline{\smash{0.079 ± 0.058}} & \textbf{0.051 ± 0.032} \\
    \bottomrule
\end{tabular}}
\end{table*}

\smallsection{Structure Preservation:}
For a more comprehensive evaluation, we evaluate whether the reconstructed hypergraphs preserve the core structural properties of the original hypergraph.
Specifically, we compare the reconstructed hypergraph $\widehat{\mathcal{H}}^{(T)}$ and the original hypergraph $\mathcal{H}^{(T)}$ based on 12 structural properties, categorized into \textit{scalar} and \textit{distributional} properties.
Scalar properties include the number of nodes and hyperedges, average node degree, average hyperedge size, simplicial closure ratio~\cite{benson2018simplicial}, hypergraph density~\cite{hu2017maintaining}, and hypergraph overlapness~\cite{lee2021hyperedges}.
Distributional properties include the distributions of node degrees, node-pair degrees, node-triple degrees, hyperedge homogeneity~\cite{lee2021hyperedges}, and singular values of the incidence matrix.
For more detailed descriptions of each structural property, refer to\cite{supple}.

For the scalar properties, we quantify the difference by the normalized difference, spec., $|x-y| / \max(x,y)$ where $x$ and $y$ are the values of the ground truth and the reconstruction, respectively, following \cite{wang2024from}.
For the distributional properties, we compare the two distributions using the Kolmogorov-Smirnov (KS) D-statistic, which quantifies the maximum difference between their cumulative distribution functions. 
For both metrics, lower values indicate better preservation of the corresponding structural property.

As shown in Table~\ref{tab:reconstruction_analysis}, \method achieves the lowest values across all scalar properties and three out of five distributional properties.
This demonstrates \method's strong capability to reconstruct hypergraphs that accurately preserve the structural properties in the original hypergraphs.

\begin{table*}[ht]
    \centering
    \caption{\small{\underline{\smash{Transfer learning performance.}} 
    Each method is trained on a source dataset and evaluated on a target dataset.
    \method achieves the highest Jaccard similarity (scaled by a factor of 100) compared to the baselines. 
    The best performance is shown in \textbf{bold}, and the second best is \underline{\smash{underlined}}.
    "OOT" indicates "out of time" (exceeding 24 hours), and "OOM" represents "out of memory" (system limit: 384 GB).
    }}
    \label{exp:transfer}
    \setlength\tabcolsep{6.8pt} 
    \renewcommand{\arraystretch}{1.2} 
    \scalebox{0.93}{ 
    \begin{tabular}{l|cccc|cc|cc}
        \toprule
        \textbf{Source Dataset} & \multicolumn{4}{c|}{\textbf{DBLP}} & \multicolumn{2}{c|}{\textbf{Eu}} & \multicolumn{2}{c}{\textbf{P.School}} \\
        \textbf{Target Dataset} & \textbf{DBLP} & \textbf{MAG-History} & \textbf{MAG-TopCS} & \textbf{MAG-Geology} & \textbf{Eu} & \textbf{Enron} & \textbf{P.School} & \textbf{H.School} \\
        \midrule
        \wangunsup     & OOT              & OOT              & \underline{84.50 $\pm$ 0.00}          & OOT              & 5.19 $\pm$ 0.00           & 13.74 $\pm$ 0.00         & 8.34 $\pm$ 0.00          & 17.00 $\pm$ 0.00 \\
        \wang-Motif     & $85.99 \pm 0.04$ & $76.73 \pm 0.52$ & OOT              & OOT              & OOM              & OOM             & OOT              & OOT \\
        \wang-Count     & \underline{86.07 $\pm$ 0.00} & \underline{93.86 $\pm$ 0.34} & $78.08 \pm 0.36$ & \underline{48.64 $\pm$ 0.31} & \underline{11.14 $\pm$ 0.27} & \underline{23.08 $\pm$ 0.08} & \underline{42.87 $\pm$ 1.66} & \underline{43.83 $\pm$ 0.28} \\
        \midrule
        \textbf{\method}   & \textbf{98.13 $\pm$ 0.06} & \textbf{97.80 $\pm$ 0.57} & \textbf{92.92 $\pm$ 0.22} & \textbf{81.66 $\pm$ 0.73} & \textbf{13.98 $\pm$ 0.23} & \textbf{27.59 $\pm$ 2.50} & \textbf{47.48 $\pm$ 0.65} & \textbf{49.52 $\pm$ 2.09} \\
        \bottomrule
    \end{tabular}}
    \vspace{5pt}
\end{table*}

\subsection{\bf Q2. Transferability}
We assess the transferability of \method by evaluating its performance in two distinct scenarios: (1) \textit{transfer setting} where \method is trained on a source dataset and then used to reconstruct a hypergraph in a different target dataset and (2) \textit{semi-supervised setting} where \method is trained on a source dataset with limited supervision.
Both settings aim to evaluate \method's adaptability and resilience under varying levels of data availability and domain shifts.

\smallsection{Transfer Learning:}
In this setting, \method is trained on source datasets and applied to target datasets within similar domains. 
As shown in Table~\ref{exp:transfer}, \method and its competitors are trained on the DBLP, Eu, P.School datasets and then used to reconstruct hypergraphs on target datasets within similar domains but with potentially different structures.
\method consistently and significantly outperforms the baselines in transfer learning across three domains. 
For example, \method achieves a 67.88\% higher Jaccard similarity compared to \wang-Count.
These results demonstrate the adaptability of \method and its strong ability to generalize across domains. 


\begin{table}[t]
    \centering
    \caption{\small{\underline{\smash{Semi-supervised learning performance.}} 
    \method maintains high reconstruction accuracy (in terms of Jaccard similarity scaled 100), even with reduced training ratios.
    The best performance is shown in \textbf{bold}, and the second best is \underline{\smash{underlined}}.
    }}\label{exp:semi}
    \resizebox{0.485\textwidth}{!}{%
    \begin{tabular}{l|ccc}
    \toprule
    \textbf{Method} & \textbf{DBLP} & \textbf{Hosts} & \textbf{Enron} \\
    \midrule
    {\bayesian} & 86.06 $\pm$ 0.01 & 53.10 $\pm$ 0.32 & 4.77 $\pm$ 0.26\\
    {\wang-Motif} & 85.99 $\pm$ 0.04 & 49.69 $\pm$ 0.59 & 14.14 $\pm$ 3.27\\
    {\wang-Count} & 86.07 $\pm$ 0.00 & 56.64 $\pm$ 0.55 & 14.36 $\pm$ 0.50\\
    \midrule
    \textbf{\method (10\%)} & 96.52 $\pm$ 0.22 & 55.42 $\pm$ 2.17 & 22.29 $\pm$ 1.78\\
    \textbf{\method (20\%)} & 97.11 $\pm$ 0.15 & 56.24 $\pm$ 1.46 & 22.45 $\pm$ 1.14\\
    \textbf{\method (50\%)} & \underline{97.88 $\pm$ 0.08} & \underline{58.17 $\pm$ 2.74} & \underline{24.10 $\pm$ 1.32} \\
    \textbf{\method (100\%)} & \textbf{98.13 $\pm$ 0.06} & \textbf{61.52 $\pm$ 2.63} & \textbf{25.06 $\pm$ 0.60} \\
    \bottomrule
    \end{tabular}%
    }
\end{table}

\smallsection{Semi-Supervised Learning:}
We evaluate \method using only 10\%, 20\%, and 50\% of the available supervision (i.e., hyperedges in the source hypergraph $\mathcal{H}^{(S)}$).
As shown in Table~\ref{exp:semi}, \method demonstrates robust performance, maintaining high Jaccard similarity even with reduced training ratios. 
Notably, DBLP and Enron, \method outperforms its baselines that are trained with full supervision, even when trained with only 10\% of the supervision.
This demonstrates the robustness of \method against limited training data.




\subsection{\bf Q3. Applicability}
\label{sec:exp:application}
We evaluate the utility of the hypergraphs reconstructed by \method in downstream tasks, specifically focusing on node clustering and link prediction. 

\smallsection{Node Clustering:}
We perform node clustering on the P.School and H.School datasets, both of which include node labels, allowing for numerical evaluation. 
Specifically, we apply spectral clustering to the projected graphs, reconstructed hypergraphs, and ground-truth hypergraphs. 
Clustering performance is measured using Normalized Mutual Information (NMI).
As shown in Table~\ref{tab:nmi_scores}, \method achieves the highest NMI scores among all reconstructed hypergraphs.
In addition, it is worthwhile to note that, ground-truth hypergraph delivers the best performance, significantly outperforming their projected graphs, implying the importance of higher-order information.


\smallsection{Node Classification:} We evaluate the effectiveness of our method on a node classification task using datasets with available node labels. Specifically, we generate node embeddings via spectral decomposition of Laplacian matrices derived from both the weighted projected graphs (obtained through clique expansion) and the original (or reconstructed) hypergraphs. These embeddings serve as input features for an MLP classifier. For a fair comparison, experiments are conducted over multiple random train/test splits and performance is quantified using Micro and Macro F1 scores.

Our experimental results, presented in Table~\ref{tab:node_classification}, demonstrate that embeddings computed from the hypergraph Laplacian consistently outperform those derived from the projected graphs. This indicates that preserving higher-order relationships within the hypergraph structure significantly enhances the discriminative power of the node representations.

\smallsection{Link Prediction:}
For the link prediction task, we utilize the same 10 datasets previously employed to evaluate reconstruction accuracy. Each edge in the graph \( G \) is paired with an equal number of randomly sampled non-edges to form a balanced dataset. This dataset is split into 90\% for training and 10\% for testing to prevent information leakage. To eliminate bias, any hyperedge in the reconstructed hypergraph \( \widehat{\mathcal{H}} \) that contains a test edge is excluded, as shared membership in a hyperedge inherently indicates a link. Test edges are removed during training and thus do not affect node embedding.

In projected graph settings, we use 
the Jaccard index, Adamic-Adar index, preferential attachment score, resource allocation score, (mean, minimum, and maximum of) node degrees, and edge weights as edge features.
In hypergraph settings, we use the same features derived from projected graphs and two extra hypergraph-specific features: the hyperedge Jaccard index\footnote{For two nodes \(u\) and \(v\), their hyperedge Jaccard index is defined as \(\frac{|HE(u)\,\cap\,HE(v)|}{|HE(u)\,\cup\,HE(v)|}\), where \(HE(u)\) is the set of hyperedges containing \(u\).} and the hyperedge size.\footnote{For each node \(u\), we let \(\overline{S}(u) = \frac{1}{|HE(u)|}\sum_{e \in HE(u)}|e|\) be the average size of hyperedges containing $u$. For two nodes $u$ and $v$, we compute the hyperedge size feature as \(\big(\min\{\overline{S}(u), \overline{S}(v)\},\,\max\{\overline{S}(u), \overline{S}(v)\}\big)\).}
We also compute link embeddings as additional features by pooling the embeddings of the two incident nodes generated by a two-layered Graph Convolutional Network (GCN) on the (projected) graph with one-hot encodings as the initial node features. The pooling operation is performed by concatenating the element-wise minimum and maximum of the node embeddings.
We apply this approach consistently across both projected graph and hypergraph settings.
To ensure fairness, all methods use the same GCN architecture for node embedding generation, with identical training configurations.

As shown in Table~\ref{tab:auc_recall_comparison}, where AUC is used as the evaluation metric, 
hypergraph-based approaches consistently outperform projected graph-based approaches. Moreover, \method leads to the highest AUC among all reconstruction methods.
Using projected graphs performs competitively on some datasets like H.School and DBLP but underperforms on others, such as Foursquare and Hosts, due to its inability to capture higher-order interactions. Overall, these results support the usefulness of hypergraphs, especially those reconstructed by \method, in leveraging higher-order information for downstream tasks.

\begin{table}[t]
    \centering
    \caption{\small{\underline{\smash{Node clustering performance.}} 
    Spectral clustering on hypergraphs reconstructed by \method achieves higher NMI than those reconstructed by the baselines. 
    The best performance is highlighted in \textbf{bold}, and the second best is \underline{\smash{underlined}}.
    }}
    \label{tab:nmi_scores}
    \setlength\tabcolsep{7.2pt}
    \scalebox{0.99}{
    \begin{tabular}{l|cc}
        \toprule
        \textbf{Input of Spectral Clustering} & \textbf{P.School} & \textbf{H.School} \\
        \midrule
        Projected graph \textbf{$\mathcal{G}$} & 0.8488 & 0.9392 \\
        \midrule
        $\widehat{\mathcal{H}}$ by \wangunsup & 0.8982 & 0.9635 \\
        $\widehat{\mathcal{H}}$ by \wang-Motif & OOT & 0.9830 \\
        $\widehat{\mathcal{H}}$ by \wang-Count & 0.9095 & 0.9874 \\
        \textbf{$\widehat{\mathcal{H}}$ by \method} & \underline{0.9234} & \textbf{0.9936} \\
        \midrule
        Original Hypergraph \textbf{$\mathcal{H}$} & \textbf{0.9255} & \textbf{0.9936} \\
        \bottomrule
    \end{tabular}}
\end{table}

\begin{table}[t]
    \centering
    \caption{\small{\underline{\smash{Node classification performance.}} 
    We evaluate micro-F1 and macro-F1 scores on the P.School and H.School datasets. 
    Hypergraphs reconstructed by \method tend to lead to higher F1 scores than those reconstructed by the baselines. 
   The best performance is highlighted in \textbf{bold}, and the second-best is \underline{\smash{underlined}}.
    }}
    \label{tab:node_classification}
    \setlength\tabcolsep{6.5pt}
    \scalebox{0.87}{
    \begin{tabular}{l|cc|cc}
        \toprule
        \multirow{2}{*}{\textbf{Input of Node Classification}} 
        & \multicolumn{2}{c|}{\textbf{Micro-F1}} 
        & \multicolumn{2}{c}{\textbf{Macro-F1}} \\
        \cmidrule(lr){2-3} \cmidrule(lr){4-5}
         & \textbf{P.School} & \textbf{H.School}
         & \textbf{P.School} & \textbf{H.School} \\
        \midrule
        Projected graph \textbf{$\mathcal{G}$}
            & 0.8380  
            & 0.8890 
            & 0.7867 
            & 0.8605 \\
        \midrule
        $\widehat{\mathcal{H}}$ by \wangunsup  
            & 0.8959  
            & 0.9720 
            & 0.8477 
            & 0.9687 \\
        $\widehat{\mathcal{H}}$ by \wang-Motif
            & OOT 
            & 0.9780 
            & OOT 
            & 0.9744 \\
        $\widehat{\mathcal{H}}$ by \wang-Count
            & 0.8760 
            & 0.9768 
            & 0.8322  
            & 0.9736 \\
        \textbf{$\widehat{\mathcal{H}}$ by \method}
            & \underline{0.9140}
            & \underline{0.9793}
            & \underline{0.8594} 
            & \underline{0.9765} \\
        \midrule
        Original Hypergraph $\mathcal{H}$ 
            & \textbf{0.9240}  
            & \textbf{0.9829} 
            & \textbf{0.8720}
            & \textbf{0.9800} \\
        \bottomrule
    \end{tabular}}
\end{table}



\begin{table*}[h]
\vspace{-2mm}
\centering
\caption{\small{\underline{\smash{Link prediction performance.}} 
Using hypergraphs reconstructed by \method tends to result in higher accuracy (in terms of AUC) compared to those reconstructed by the baselines.
Average AUC scores across five random seeds are reported with standard deviation. 
The best performance is shown in \textbf{bold}, and the second best is \underline{\smash{underlined}}.
"OOT" indicates "out of time" (exceeding 24 hours).
}}
\label{tab:auc_recall_comparison}
\setlength\tabcolsep{2.7pt}
\scalebox{0.845}{
\begin{tabular}{l|cccccccccc|c}
\toprule
\textbf{Input} & \textbf{Enron} & \textbf{P.School} & \textbf{H.School} & \textbf{Crime} & \textbf{Hosts} & \textbf{Directors} & \textbf{Foursquare} & \textbf{DBLP} & \textbf{Eu} & \textbf{MAG-TopCS} & \textbf{Avg. Rank} \\
\midrule
{Projected graph $\mathcal{G}$} & \underline{83.94 ± 2.99}& 90.76 ± 0.37 & \textbf{92.51 ± 1.37} & 78.32 ± 1.11 & 98.61 ± 0.06 & 81.90 ± 3.10 & 99.44 ± 0.02 & \textbf{96.54 ± 0.09} & \textbf{97.46 ± 0.06} & 94.86 ± 0.22 & 3.90 ± 2.33 \\
\midrule
{$\widehat{\mathcal{H}}$ by \wangunsup} & 80.07 ± 4.26 & 91.21 ± 0.31 & 90.08 ± 1.40 & 81.38 ± 0.80 & 98.75 ± 0.05 & 82.59 ± 3.80 & \textbf{99.60 ± 0.02} & OOT & 97.18 ± 0.10 & 95.05 ± 0.22 & 3.50 ± 1.51 \\
{$\widehat{\mathcal{H}}$ by \wang-Motif} & 78.65 ± 3.73 & OOT & 89.92 ± 1.64 & 81.54 ± 0.94 & \underline{98.76 ± 0.03} & \underline{82.59 ± 3.80} & 99.57 ± 0.02 & 96.40 ± 0.11 & OOT & 94.94 ± 0.14 & 4.40 ± 1.58 \\
{$\widehat{\mathcal{H}}$ by \wang-Count} & 77.76 ± 3.09 & 91.04 ± 0.77 & 90.16 ± 1.57 & 81.54 ± 0.94 & \textbf{98.77 ± 0.11} & 82.58 ± 3.80 & 99.57 ± 0.01 & 96.40 ± 0.04 & 95.82 ± 0.61 & 94.95 ± 0.19 & 3.80 ± 1.32 \\
\textbf{$\widehat{\mathcal{H}}$ by \method} & 79.75 ± 4.74 & \underline{91.72 ± 0.22} & 90.21 ± 1.51 & \textbf{81.60 ± 1.00} & 98.68 ± 0.06 & \textbf{82.75 ± 3.71} & \textbf{99.60 ± 0.02} & \underline{96.46 ± 0.06} & 97.19 ± 0.09 & \textbf{95.10 ± 0.17} & \textbf{2.30 ± 1.42} \\
\midrule
{Original Hypergraph $\mathcal{H}$} & \textbf{84.55 ± 3.20} & \textbf{92.16 ± 0.19} & \underline{90.35 ± 1.18} & \underline{81.59 ± 0.82} & 98.70 ± 0.06 & 82.58 ± 3.80 & \textbf{99.60 ± 0.02} & 96.39 ± 0.06 & \underline{97.26 ± 0.13} & \underline{95.06 ± 0.14} & \underline{2.40 ± 1.43} \\
\bottomrule
\end{tabular}}
\vspace{5pt}
\end{table*}

\subsection{\bf Q4. Effectiveness}
We evaluate the effectiveness of each component, the importance of multiplicity-aware features, and the robustness of \method to changes in its hyperparameters through ablation studies and sensitivity analysis.

\smallsection{Effects of Multiplicity-Aware Classifier:}\label{sec:exp:multiplicity}
We evaluate the impact of removing multiplicity-aware features from \method by replacing its features with those from \wang-Count~\cite{wang2024from}, a baseline without multiplicity-aware features. This modified version, denoted as \method-M, consistently underperforms compared to \method across all datasets, as shown in Tables~\ref{tab:jaccard_comparison} and~\ref{tab:mjaccard_comparison}. This highlights the importance of multiplicity-aware features for reconstruction accuracy.


Nevertheless, \method-M outperforms \wang-Count on most datasets (except Hosts), demonstrating that \method's candidate search strategy remains more effective even without multiplicity-aware features. This study supports the effectiveness of two key components of \method: its candidate search mechanism and multiplicity-aware clique features.

\smallsection{Effects of Filtering:}
We assess the impact of filtering in \method by comparing its performance with a version without filtering, denoted as \method-F. As shown in Tables~\ref{tab:jaccard_comparison} and~\ref{tab:mjaccard_comparison}, \method consistently outperforms \method-F across all datasets, highlighting the importance of filtering for improved reconstruction accuracy. Even without filtering, \method-F surpasses other baselines, as size-2 hyperedges, guaranteed by theory, remain in the candidate space. Over iterations, these candidates are assigned high prediction scores and eventually reconstructed as hyperedges.

\smallsection{Effects of Bidirectional Search:}\label{sec:exp:bidirectional}
We assess the impact of bidirectional search in \method by modifying the reconstruction process to skip sub-clique evaluation for low-scoring candidate cliques, creating a variant called \method-B. As shown in Tables~\ref{tab:jaccard_comparison} and~\ref{tab:mjaccard_comparison}, \method-B shows varied performance across datasets. For example, it performs significantly worse than \wang-Count on P.School and H.School, emphasizing the importance of bidirectional search in these cases. However, on other datasets, \method-B outperforms all baselines and even surpasses \method on MAG-TopCS.

This unexpected result on MAG-TopCS suggests that skipping sub-clique evaluation can preserve large candidate cliques that align better with true hyperedges in certain datasets, indicating that bidirectional search, while typically beneficial, may over-prune candidates in specific cases.

\begin{figure}
    \centering
    \includegraphics[width=0.59\linewidth]{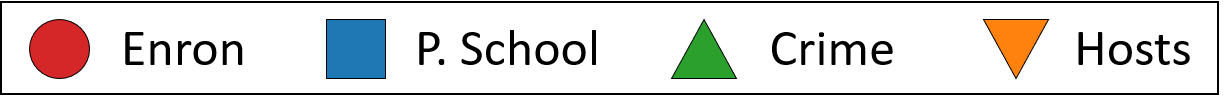}\\
    \begin{subfigure}[b]{.164\textwidth}
        \centering
        \includegraphics[width=0.99\linewidth]{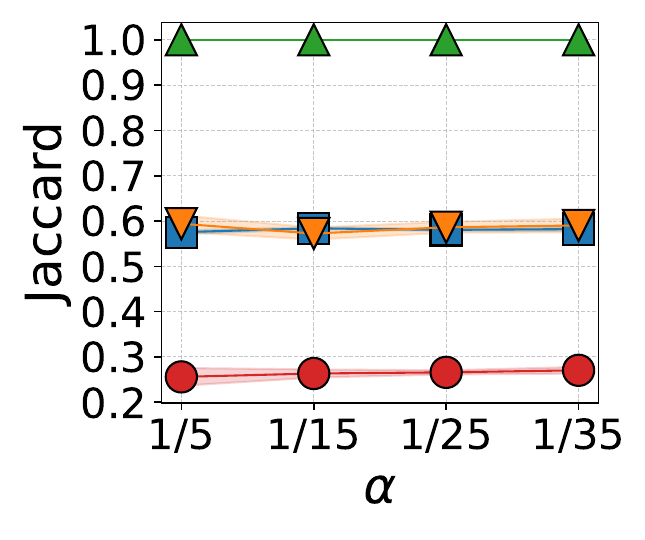}
    \end{subfigure}
    \hspace{-10pt}
    \begin{subfigure}[b]{.164\textwidth}
        \centering
        \includegraphics[width=0.99\linewidth]{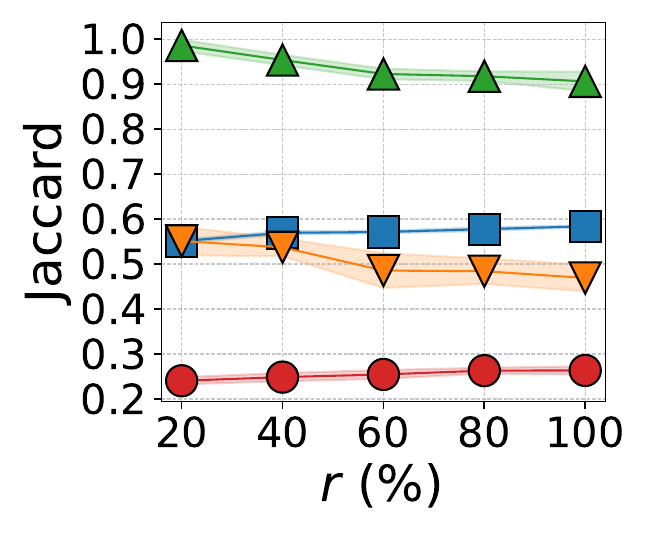}
    \end{subfigure}
    \hspace{-10pt}
    \begin{subfigure}[b]{.164\textwidth}
        \centering
        \includegraphics[width=0.99\linewidth]{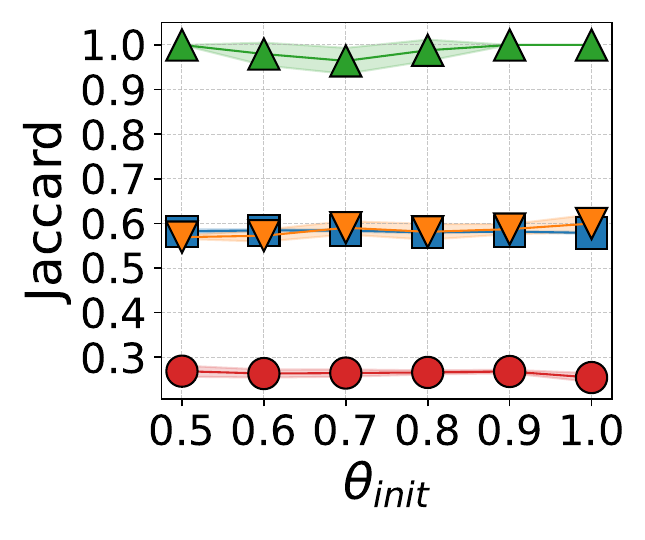}
    \end{subfigure}
    \begin{subfigure}[b]{.164\textwidth}
        \centering
        \includegraphics[width=0.99\linewidth]{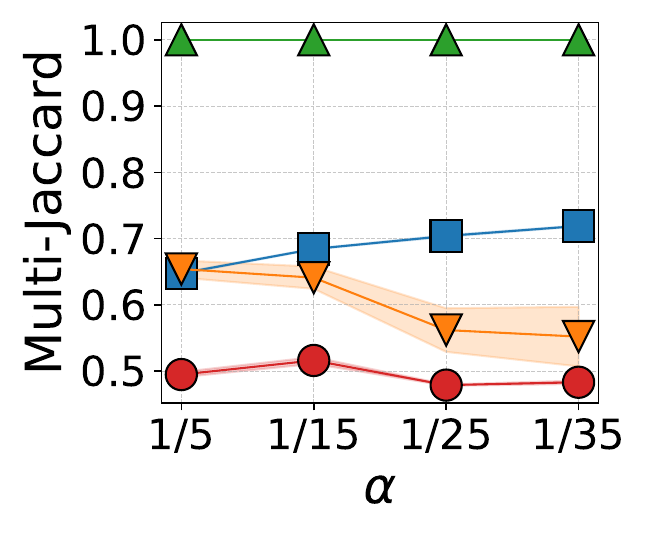}
    \end{subfigure}
    \hspace{-10pt}
    \begin{subfigure}[b]{.164\textwidth}
        \centering
        \includegraphics[width=0.99\linewidth]{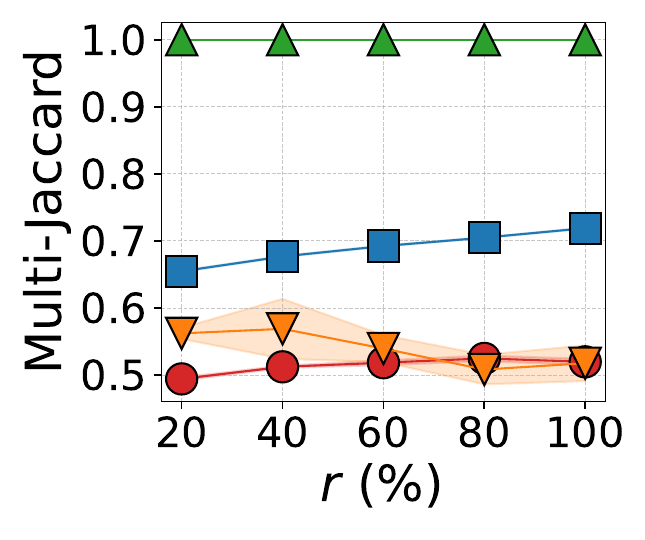}
    \end{subfigure}
    \hspace{-10pt}
    \begin{subfigure}[b]{.164\textwidth}
        \centering
        \includegraphics[width=0.99\linewidth]{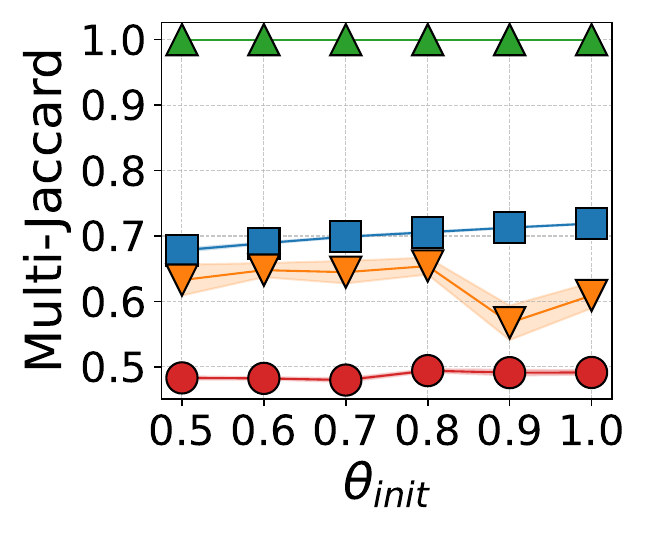}
    \end{subfigure}
    \vspace{-9pt}
    \caption{\small{\underline{\smash{Hyeprparameter sensitivity analysis.}} 
    \method is robust to variations in hyperparameters: $\alpha$, $r$, and $\theta_\text{init}$, in both multiplicity-reduced setting (above) and multiplicity-preserved setting (below).
    }}
    \label{exp:sensitivity}
\end{figure}


\smallsection{Parameter Sensitivity:}
We assess the sensitivity of \method to its key hyperparameters, including the initial classification threshold (\(\theta_{\text{init}}\)), the negative prediction processing ratio (\(r\)), and the threshold adjust ratio (\(\alpha\)). 
Specifically, hyperparameters are explored from the ranges: \(\theta_t = [0.5, 0.6, \dots, 1.0]\), \(r = [5, 10, \dots, 100]\), and \(\alpha = [\frac{1}{5}, \frac{1}{15}, \frac{1}{25}, \frac{1}{35}]\). 
As shown in Fig.~\ref{exp:sensitivity}, \method maintains robust performance across most datasets, with consistent reconstruction accuracy under varying settings.
However, in the Hosts dataset, changes in \(\alpha\) and \(r\) cause some fluctuations in the Jaccard and multi-Jaccard similarities, suggesting that datasets with specific structural characteristics may require more careful hyperparameter tuning for optimal results.

\subsection{\bf Q5. Scalability} \label{sec:exp:runningtime}
\begin{figure}[t]
    \centering
    \includegraphics[width=0.45\linewidth]{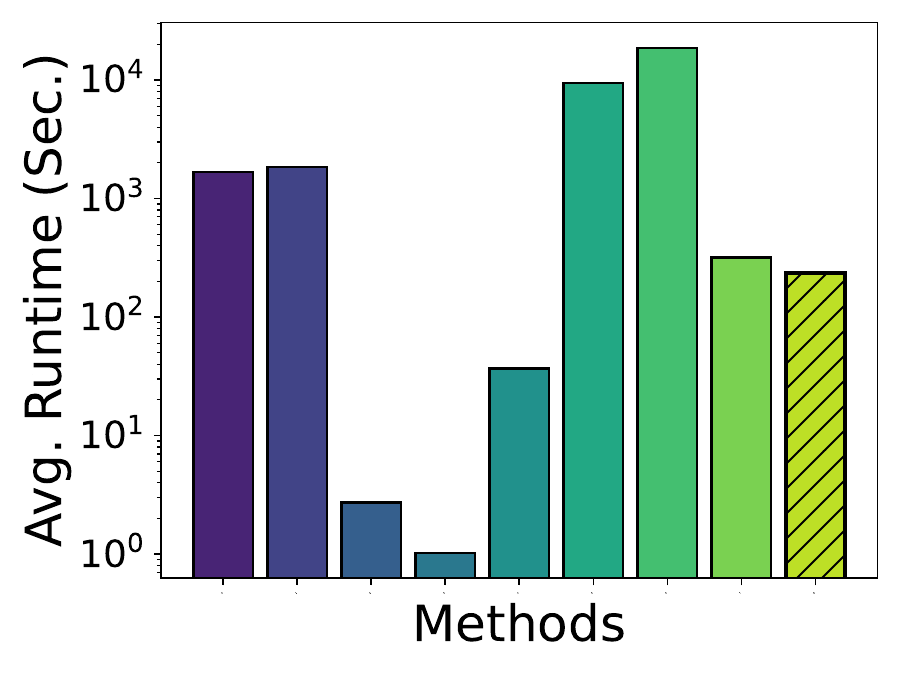}
    \hspace{2pt}
    \raisebox{0.32\height}{%
        \includegraphics[width=0.45\linewidth]{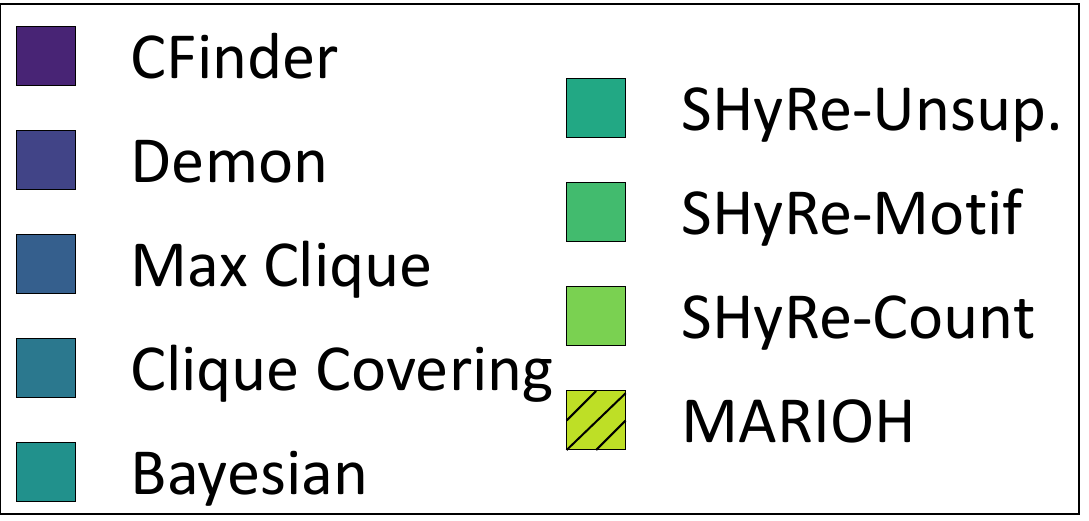}
    }
    \vspace{-3mm}
    \caption{\small{\underline{\smash{Average runtime of \method and competitors.}} 
    While \method is slower than basic baselines, it takes less time on average than recent advanced methods (i.e.,  \wang-Motif and \wang-Count).
    For runtimes on individual datasets, see Fig.~\ref{exp:run_compare} and the online appendix~\cite{supple}.}}
    \label{exp:runtime}
\end{figure}

We compare the runtime of \method with competing methods. As shown in Fig.~\ref{exp:runtime}, \method is faster than community-based methods like CFinder and Demon. While clique decomposition methods run the fastest, their reconstruction accuracy is much lower (Tables~\ref{tab:jaccard_comparison} and \ref{tab:mjaccard_comparison}). Among hypergraph reconstruction methods, \bayesian is the fastest, followed closely by \method. \wang-Count is slightly slower than \method on average (see detailed comparisons below), while \wang-Motif and \wangunsup show significantly longer execution times. 

\begin{figure}[t]
    \vspace{-3mm}
    \centering
    \hspace{6mm}
    \includegraphics[width=0.85\linewidth]{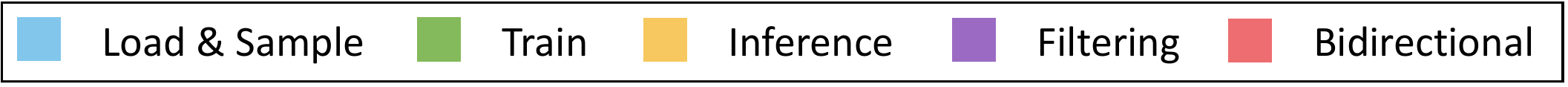}\\
    \includegraphics[width=\linewidth]{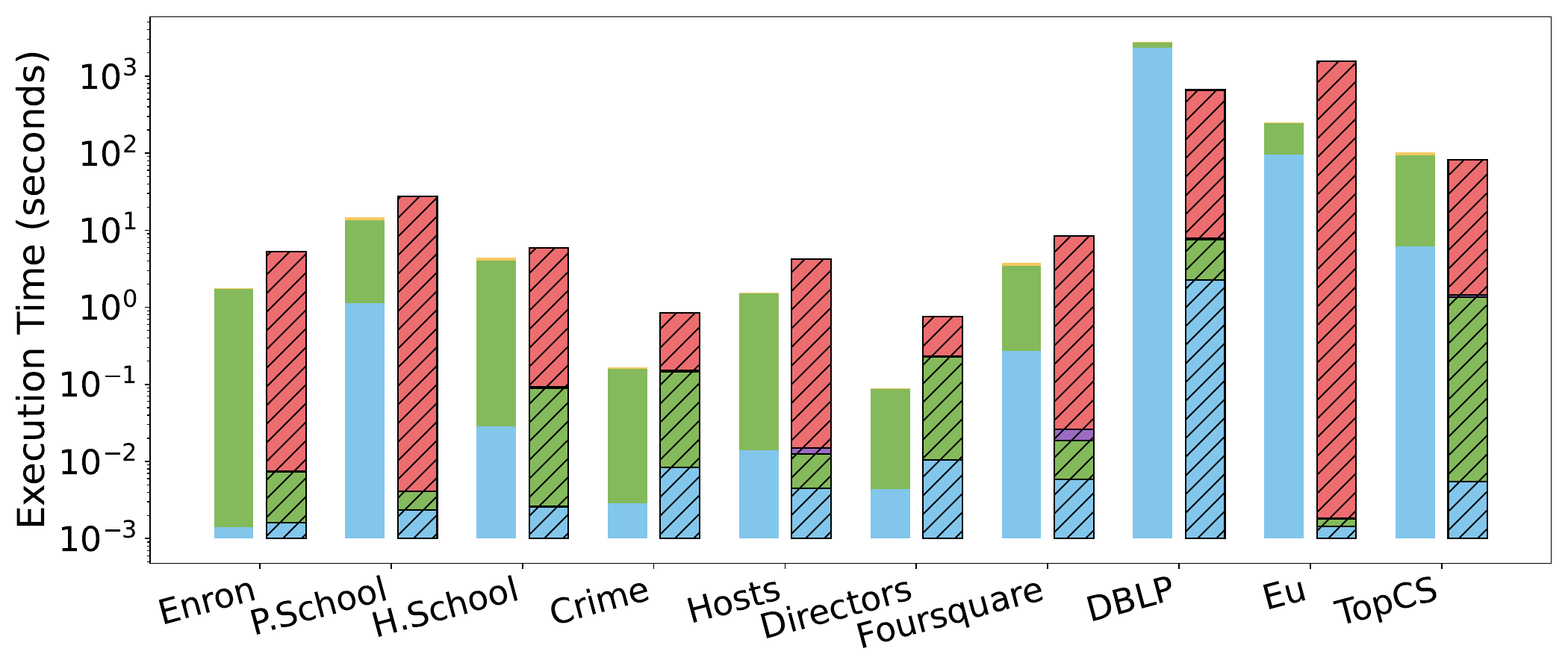}
    \vspace{-6mm}
    \caption{\small{\underline{\smash{Runtime comparison of \method and \wang-Count.}}
    The left bar (solid color) for each dataset corresponds to \wang-Count, while the right bar (hatched) corresponds to \method. 
    Each segment within a bar represents the proportion of the execution time of the corresponding step relative to the total execution time.
    The inference step in \wang-Count corresponds to the filtering and bidirectional-search steps in \method.
    }}
    \label{exp:run_compare}
\end{figure}

We provide a detailed comparison of \method and \wang-Count across all 10 real-world hypergraph datasets.
Regarding reconstruction accuracy, as summarized in Tables~\ref{tab:jaccard_comparison} and \ref{tab:mjaccard_comparison}, \method consistently achieves higher Jaccard and multiset Jaccard similarities across all the datasets. This improvement stems from the \texttt{Filtering} and \texttt{Bidirectional Search} steps in \method, which iteratively refine and discover new candidate cliques, whereas \wang-Count classifies only a fixed set of candidates. 
However, these steps come at the cost of speed, as shown in Fig.~\ref{exp:run_compare}, where \method generally takes more time than \wang-Count on most datasets, except for the largest DBLP dataset.

\begin{figure}[t]
    \vspace{-3mm}
    \centering
    \captionsetup[subfigure]{aboveskip=1pt, belowskip=1pt} 
    \begin{subfigure}[b]{.225\textwidth}
        \centering
        \includegraphics[width=\linewidth]{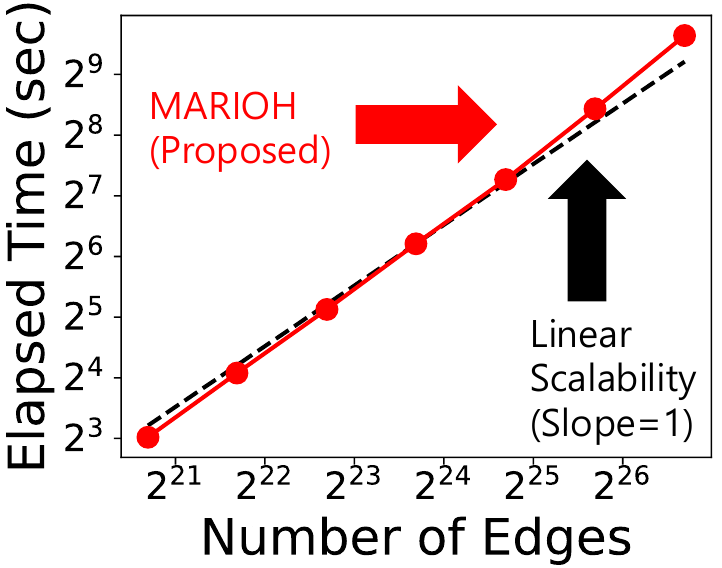}
        \caption{\centering\small{Filtering Step}}\label{exp:filtering_scale}
    \end{subfigure}
    \hspace{10pt}
    \begin{subfigure}[b]{.225\textwidth}
        \centering
        \includegraphics[width=\linewidth]{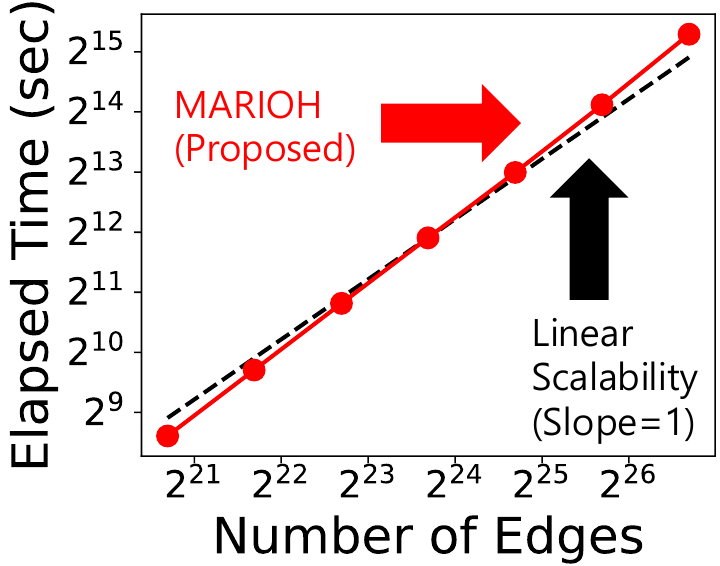}
        \caption{\centering\small{Bidirectional Search Step}}\label{exp:bd_scale}
    \end{subfigure}
    \vspace{-3pt}
    \caption{\label{exp:scalability2} \small{\underline{\smash{Scalability analysis.}} 
    \method scales nearly linearly with the data size (spec., the number of edges in the projected graph). 
    }}
\end{figure}

Additionally, we evaluate the scalability of \method on datasets of varying sizes, which are generated using HyperCL \cite{lee2021hyperedges} with DBLP dataset statistics as input.
Note that for all these datasets, the original DBLP dataset is used for training, and thus the training time is independent of the dataset size.
As shown in Fig.~\ref{exp:scalability2}, the running times of both the \texttt{Filtering} and \texttt{Bidirectional Search} steps of \method scale nearly linearly (with a slope close to 1 on a log-log scale) as the number of edges in the input graph increases.
This result demonstrates that \method is practically useful for large-scale datasets.

\smallsection{Extra Experiments:}
In the online appendix \cite{supple}, we present analyses of (1) feature importance, (2) storage savings through hypergraph reconstruction, and (3) runtimes on individual datasets, in addition to more case studies.

\section{Related Work}
\label{sec:related}



In this section, we review related studies on hypergraph reconstruction. Then, we discuss the broader connections of our work to diverse database studies.


\smallsection{Hypergraph Reconstruction:} Hypergraph reconstruction methods seek to recover higher-order interactions from projected graphs. Young et al.\cite{young2021hypergraph} proposed a Bayesian approach (\bayesian) that reconstructs hypergraphs using a Bayesian generative model, emphasizing the principle of parsimony to infer the simplest hypergraph that explains the observed pairwise data. It calculates the posterior probability of a hypergraph and estimates structures through Markov Chain Monte Carlo sampling. However, this assumption of parsimony often falls short in practice\cite{wang2024from}. Wang and Kleinberg \cite{wang2024from} introduced \wang, a supervised method that reconstructs hypergraphs by sampling cliques from the projected graph based on a statistical distribution \(\rho(n,k)\), which captures patterns of hyperedges within maximal cliques. A classifier is trained to distinguish between hyperedges and non-hyperedges. Specifically, \wang-Count uses basic structural features, while \wang-Motif incorporates subgraph patterns such as triangle and square motifs as additional features for classification. However, its reliance on sampling leads to missed hyperedges (false negatives) and ignores edge multiplicity. To address edge multiplicity, they also proposed an unsupervised method (\wangunsup) that reconstructs hypergraphs by iteratively selecting maximal cliques as candidates for hyperedges. The cliques are ranked based on their size and average edge multiplicity, preferring larger cliques with lower average edge multiplicities. Once selected, a clique is converted into a hyperedge, and the corresponding edge multiplicities in the projected graph are reduced until all edge multiplicities are 0. However, this approach struggles with scalability, as it involves repeated intensive computations, and its reconstruction accuracy sometimes falls behind \wang, revealing limitations in both existing methods for effective hypergraph reconstruction.

\smallsection{Connections to Database Research:} Database research addresses data enhancement tasks, including data integration\cite{doan2012principles, lenzerini2002data, rahm2001survey}, record linkage\cite{christen2012data, bilenko2006adaptive, elmagarmid2006duplicate}, and data cleaning\cite{rahm2000data, chu2016data}. In these tasks, uncovering associations among data elements is crucial, even when the available data is incomplete or simplified. For instance, data integration tackles the challenge of merging disparate data sources by identifying underlying relationships. In record linkage, detecting connections is essential for accurately matching records referring to the same entity. Similarly, in data cleaning, recovering relationships is important to improve data consistency and quality. 
Hypergraph reconstruction can also be considered a data enhancement task, where restoring higher-order relationships enhances the data quality for subsequent analysis.
Moreover, because hypergraph representations often offer storage savings over graph representations (see the online appendix~\cite{supple} for a detailed storage savings analysis), Hypergraph restoration can also be seen as a data compression task, particularly related to lossy graph compression~\cite{qiao2017subgraph, fan2012query, navlakha2008graph, xu2024improving}.

\section{Conclusion}
\label{sec:conclusion}

This work presents \method, a supervised, multiplicity-aware hypergraph reconstruction method that addresses the challenges of recovering hypergraphs from projected graphs. 
In essence, \method incorporates a bidirectional greedy search and multiplicity-aware features to enhance reconstruction efficiency and accuracy, leading to the following strengths:
\begin{itemize}[leftmargin=*]
    \item \textbf{Accurate Recovery:} Achieving up to 74.51\% higher accuracy than existing methods, \method effectively reconstructs hypergraphs with and without hyperedge multiplicity.
    \item \textbf{Transferability:} Generalizing effectively across datasets within the same domain, \method maintains robustness and adaptability without requiring retraining.
    \item \textbf{Applications:} Improving performance in downstream tasks such as link prediction and node clustering, \method-reconstructed hypergraphs demonstrate the benefits of recovering higher-order structures.
\end{itemize}
These properties enable \method to scale to large datasets and remain versatile across diverse datasets.

\smallsection{Reproducibility:} The code and data used in the paper can be found at  \url{https://github.com/KyuhanLee/MARIOH}.

{\small  \smallsection{Acknowledgements:}
This work was partly supported by the National Research Foundation of Korea (NRF) grant funded
by the Korea government (MSIT) (No. RS-2024-00406985, 30\%).
This work was partly supported by Institute of Information \& Communications Technology Planning \& Evaluation (IITP) grant funded by the Korea government (MSIT) (No. 2022-0-00871 / RS-2022-II220871, Development of AI Autonomy and Knowledge Enhancement for AI Agent Collaboration, 30\%) 
(No. RS-2024-00438638, EntireDB2AI: Foundations and Software for Comprehensive Deep Representation Learning and Prediction on Entire Relational Databases, 30\%) (No. RS-2019-II190075, Artificial Intelligence Graduate School Program (KAIST), 10\%).}


\bibliographystyle{IEEEtran}
\bibliography{reference}


\end{document}